%% file: main.tex
\documentclass[11pt,a4paper]{article}
\usepackage[left=2.35cm,top=3cm,bottom=3cm,right=2.3cm]{geometry}
\usepackage{authblk}
\usepackage{amsmath}
\usepackage{tikz}
\usepackage{float}
\usepackage{cite}
\usepackage[caption=false]{subfig}
\usepackage[nolist,nohyperlinks]{acronym}
\usepackage[colorlinks=true
,urlcolor=blue
,anchorcolor=blue
,citecolor=blue
,filecolor=blue
,linkcolor=blue
,menucolor=blue
,linktocpage=true
,pdfproducer=medialab
]{hyperref}
\input{orcidlink}
\usepackage{slashed}

\def\D{{\rm d}}
\def\th{{\rm th}}
\def\sth{{4m_\pi^2}}
\def\dispint{\int_\sth^\infty}
\def\F{\mathcal{F}}
\def\Fpi{F_\pi^V}
\def\Pih{\Pi_{h}}
\def\Pires{\Pi_{\rm res}}
\def\io{i\delta}
\def\thavg{\theta_{\rm avg}}
\def\eepp{$ ee\to \pi \pi$}
\def\eeppg{$ ee\to \pi \pi \gamma$}

\def\dpone{{s_1}}
\def\dponesq{{s^2_1}}
\def\dptwo{{s_2}}

\def\mcmule{{{\sc McMule}}}
\newcommand{\OpenLoops}{{\rmfamily\scshape OpenLoops}}
\newcommand*\pFqskip{8mu}
\catcode`,\active
\newcommand*\pFq{\begingroup
        \catcode`\,\active
        \def ,{\mskip\pFqskip\relax}%
        \dopFq
}
\catcode`\,12
\def\dopFq#1#2#3#4#5{%
        {}_{#1}F_{#2}\biggl[\genfrac..{0pt}{}{#3}{#4};#5\biggr]%
        \endgroup
}

\usetikzlibrary{decorations.markings,decorations.pathmorphing,positioning,decorations.pathreplacing,shapes,calc}
\tikzset{
    photon/.style={decorate, decoration={snake,amplitude=1pt,segment length=6pt}},
    zigzag/.style={decorate, decoration=zigzag},
}

\renewcommand{\Re}{\mathrm{Re}}
\renewcommand{\Im}{\mathrm{Im}}

\hypersetup{pdftitle={Disperon QED},pdfauthor={
Yizhou Fang,
Sophie Kollatzsch,
Marco Rocco,
Adrian Signer,
Yannick Ulrich,
Max Zoller}}

\title{Disperon QED}
\date{}
\author[1]{Yizhou~Fang\,\orcidlink{0009-0003-3568-9661}\,}
\author[1,2]{Sophie~Kollatzsch\,\orcidlink{0000-0002-8560-1619}\,}
\author[3]{Marco~Rocco\,\orcidlink{0000-0002-2561-1209}\,}
\author[1,2]{Adrian~Signer\,\orcidlink{0000-0001-8488-7400}\,}
\author[4]{\\ Yannick~Ulrich\,\orcidlink{0000-0002-9947-3064}\,}
\author[1,2]{Max~Zoller\,\orcidlink{0000-0001-9233-7951}\,}
\affil[1]{PSI Center for Neutron and Muon Sciences, 5232 Villigen PSI, Switzerland}
\affil[2]{Physik-Institut, Universit\"at Z\"urich, 8057 Z\"urich, Switzerland}
\affil[3]{Universit\`a degli Studi di Torino \& INFN, 10125 Torino, Italy}
\affil[4]{University of Liverpool, Liverpool L69 3BX, U.K.}

\begin{document}

\include{figures/tikz-diags}
\definecolor{bluegray}{RGB}{140, 170, 250}

\makeatletter
    \newacro{DET}{disperon effective theory}
    \newacro{EFT}{effective field theory}
    \newacro{sQED}{scalar QED}
    \newacro{FsQED}{form-factor scalar QED}
    \newacro{FxsQED}[F$\times$sQED]{form-factor times scalar QED}
    \newacro{GVMD}{generalised vector-meson dominance}
    \newacro{HVP}{hadronic vacuum polarisation}
    \newacro{MC}{Monte Carlo}
    \newacro{MoR}{method of regions}
    \newacro{LO}{leading order}
    \newacro{NLO}{next-to-leading order\def\AC@acronyms@NNLO{@<>@<>@}\def\AC@acronyms@LO{@<>@<>@}\def\AC@acronyms@NNNLO{@<>@<>@}}
    \newacro{NNLO}{next-to-next-to-leading order\def\AC@acronyms@NNNLO{@<>@<>@}\def\AC@acronyms@NLO{@<>@<>@}\def\AC@acronyms@LO{@<>@<>@}}
    \newacro{NNNLO}[N$^3$LO]{next-to-next-to-next-to-leading order\def\AC@acronyms@NLO{@<>@<>@}\def\AC@acronyms@LO{@<>@<>@}\def\AC@acronyms@NNLO{@<>@<>@}}
    \newacro{VFF}{vector form factor}
    \newacro{VP}{vacuum polarisation}
    \newacro{dp}[{\tt dp}]{double precision}
    \newacro{qp}[{\tt qp}]{quadruple precision}
    \newacro{IR}{infrared}
    \newacro{UV}{ultraviolet}
    \newacro{ISC}{initial-state corrections}
    \newacro{FSC}{final-state corrections}
    \newacro{CT}{counterterm}
    \newacro{TPE}{two-photon exchange}
    \newacro{FKSl}[FKS$^\ell$]{}
    \newacro{NTS}{next-to-soft}
    \acrodefplural{CT}{counterterms}
\makeatother
\acused{FKSl}

\begin{titlepage}
\clearpage\maketitle 
\thispagestyle{empty}

\begin{abstract}\noindent
We present disperon QED, a method to deal with data input in loop processes in Monte Carlo codes.
It relies on dispersion relations, automated tools such as \OpenLoops, effective field theory methods and a threshold subtraction.
We motivate this method and apply it to the process \eepp{} in \mcmule{} to deal with hadronic vacuum polarisation insertions in two-loop contributions as well as the vector form factor of the pion within the form-factor \acl{sQED} approximation.
The generality of this method for more complicated processes is emphasised.
\end{abstract}
\end{titlepage}

\section{Introduction} \label{sec:introdution}

Low-energy scattering processes play a vital role in addressing many
current fundamental questions in particle physics. There is a rich
experimental program for electron-positron scattering into hadronic or
leptonic final states and for lepton-proton scattering at energies well below the electroweak scale, to name two of
the most active fields. In both cases, the situation is currently
rather confusing. If $e^+\,e^-$ scattering is used to extract the
contribution of the \ac{HVP} to the
anomalous magnetic moment of the muon~\cite{Aliberti:2025beg}, there
is a considerable tension between various experimental
results. Furthermore, most of them seem to be in conflict with recent
lattice calculations~\cite{Borsanyi:2020mff,RBC:2023pvn,Djukanovic:2024cmq}. If lepton-proton scattering is used to
extract the proton form factor at various values of the momentum
transfer, there are also tensions between different experiments and
methods used, see~\cite{Afanasev:2023gev} for a review. Furthermore, most
scattering experiments lead to a proton radius that is in conflict
with the most precise spectroscopy result of muonic
hydrogen~\cite{Pohl:2010zza}.

Precise Monte Carlo tools for such processes are crucial to
investigate possible reasons for these discrepancies. Triggered by
these observations, renewed efforts and initiatives have been started
to improve these tools~\cite{Aliberti:2024fpq, Afanasev:2023gev} and exploit the
computational advances of recent years. Regarding pure QED corrections
with pointlike constituents, there has indeed been substantial
progress. Currently, leptonic $2\to{2}$ processes have been computed at
\ac{NNLO} in QED~\cite{Bucoveanu:2018soy,Banerjee:2020rww,CarloniCalame:2020yoz,Banerjee:2021mty,Banerjee:2021qvi,Budassi:2021twh,Broggio:2022htr,Engel:2023arz} as well as in
all-order resummation improved \ac{NLO} calculations in a fully differential way~\cite{Balossini:2006wc, Balossini:2008xr,Budassi:2024whw, Price:2025fiu}. Electroweak \ac{NLO} corrections are also available and are combined with QED corrections into Monte Carlo codes~\cite{Alacevich:2018vez, Arbuzov:2022mij,Kollatzsch:2022bqa,Kollatzsch:2025pnp}. For $2\to{3}$ processes, progress towards
the same perturbative order is ongoing~\cite{Fadin:2023phc,Badger:2023xtl,Badger:2023mgf,PetitRosas:2025xhm}  and likely to lead to concrete result in the
near future. Improved theory for $2\to{3}$ processes is urgently needed, since radiative processes with an
additional photon in the final state play a decisive role in the
experimental extraction of the \ac{HVP}. 

With increasing precision in perturbative calculations for low-energy processes, the question of
how to include non-perturbative contributions becomes more
urgent. Indeed, all processes mentioned above will at some point
depend on insertions of \ac{HVP} or hadronic light-by-light contributions. For processes with external hadrons $h$, additional hadronic matrix elements such as the Compton tensor $\gamma^*\,\gamma^*\to h\,h$ need to be embedded into amplitudes. The standard procedure to do this is to decompose the hadronic matrix elements into Lorentz structures and determine their scalar coefficients numerically from data~\cite{Hoferichter:2013ama, Colangelo:2015ama, Hoferichter:2019nlq}. Once the coefficients are known, the question arises how to combine the hadronic part with the remaining amplitude. 
As long as the hadronic matrix elements only enter tree-level amplitudes, they
can be included through simple multiplications.
However, if they appear within loops, the momentum dependence of the matrix elements results in a much more complicated situation. The resulting loop integrations cannot be done with standard tools.

Here we focus on insertions of \ac{HVP} and form factors into loop calculations, taking advantage of the fact that
light-by-light contributions only appear beyond NNLO for the processes
considered here. Regarding form-factor insertions, it has been shown that using the \ac{sQED} vertex multiplied by the \ac{VFF} for photon-pion interactions corresponds to the inclusion of the pion-pole terms in the pion Compton tensor~\cite{Colangelo:2015ama}.
While this correspondence does not hold in the case of protons or with more than two photons attached to the pions, form-factor improved QED is a meaningful first step towards a more reliable description of structure-dependent corrections in low-energy processes with external hadrons~\cite{Mo:1968cg,Maximon:2000hm,Aliberti:2024fpq,Kaziukenas:2025gggpp}. 

A non-trivial inclusion of \ac{HVP} effects for scattering processes
appears only at \ac{NNLO} where it is similar to the \ac{HVP} corrections to an anomalous magnetic moment.
Recent examples include the evaluation of the HVP insertions
for muon-electron scattering at \ac{NNLO} through the
dispersive~\cite{Fael:2019nsf} or the hyperspherical
approach~\cite{Fael:2018dmz}. Based on these techniques, several fully
differential \ac{NNLO} calculations have been carried out with the
inclusion of \ac{HVP} effects~\cite{Banerjee:2021qvi,Broggio:2022htr,Budassi:2021twh}. For the anomalous
magnetic moment, even higher orders have been
computed~\cite{Balzani:2021del}. The applied techniques do not rely on
a particular functional form of the \ac{HVP}. Thus, they can be easily
adapted if a more precise non-perturbative input becomes available.

The situation regarding inclusion of form factors in loop
contributions to fully differential observables is less advanced. So
far, problems of this nature have been solved on a case by case basis.
Contributions with a pion \ac{VFF} within a loop for
\eepp{} have been computed in an
analytic way at one loop through the \ac{GVMD}~\cite{Sakurai:1972wk,Ignatov:2022iou}. Similarly, the inclusion of the proton
form factor in \ac{TPE} for elastic lepton-proton scattering
typically relies on soft-photon approximations~\cite{Mo:1968cg, Maximon:2000hm, Gramolin:2014pva, Gerasimov:2015aoa} or particular analytic representations such as a
dipole form factor~\cite{Bystritskiy:2007hw, Engel:2023arz}. However, these methods become increasingly tedious for more complicated final states. Furthermore, they often are not amenable to generic forms of the hadronic input, but rather rely on particular parametrisations thereof. \ac{TPE} contributions have also been evaluated without model dependence in an \ac{EFT} framework~\cite{Choudhary:2023rsz, Dye:2016uep, Dye:2018rgg} or through dispersion relations, e.g.~\cite{Tomalak:2016vbf, Tomalak:2017shs, Ahmed:2020uso}. Reviews focusing on TPE in lepton-proton scattering are provided in~\cite{Carlson:2007sp, Arrington:2011dn, Afanasev:2017gsk, Borisyuk:2019gym}.

In order to deal with a generic numerical input, we need a recipe that brings the dependence on the loop momentum $k$ into a form that is compatible with standard loop calculations.
The most common approach is to express the numerical function $F(k^2)$ through a (once-subtracted) dispersion relation
\begin{align}
\frac{F(k^2)}{k^2} = \frac{F(0)}{k^2} - \frac1{\pi} \int_{s_\text{thr}}^\infty \frac{\D \dpone}{\dpone} \frac{\Im\,F(\dpone)}{k^2-\dpone}\, ,
    \label{eq:disp}
\end{align}
solve the loop integral analytically and subsequently integrate numerically over the dispersion parameter~\cite{Cabibbo:1961sz}. At the price of an additional integration over the
dispersive parameter $\dpone$ starting from the pair-production threshold $s_\text{thr}$, this allows to reduce the expressions to
standard loop integrals.

In the context of \ac{VFF} insertions for the process
\eepp{}, such a procedure 
was used in~\cite{Colangelo:2022lzg} to compute the forward-backward asymmetry.  The obtained results agree well with the \ac{GVMD}
calculation~\cite{Ignatov:2022iou} and with data~\cite{CMD-3:2023alj,CMD-3:2023rfe}. These
methods were also combined with a parton shower in the
Monte Carlo code BabaYaga@NLO~\cite{Budassi:2024whw}. The results clearly
show that an external multiplication of the box diagrams with a form
factor~\cite{Hoefer:2001mx,Tracz:2018,Arbuzov:2020foj} -- a procedure dubbed as \ac{FxsQED}
in~\cite{Aliberti:2024fpq} -- gives poor agreement with experimental
data, whereas the \ac{FsQED} approach~\cite{Colangelo:2014dfa,Colangelo:2015ama,Colangelo:2022lzg} based on the dispersion relation~\eqref{eq:disp} with \ac{VFF} in the loop leads to
substantial improvements.

In view of the importance of radiative processes such as
\eeppg{}, it is desirable to extend the \ac{FsQED}
approach to $2\to{3}$ or even more complicated processes.  So far,
this process has only been considered in the \ac{FxsQED}
approach~\cite{Czyz:2003ue,Tracz:2018,Campanario:2019mjh}.
While \ac{FsQED} will not capture all
non-perturbative effects, it constitutes a major improvement, in
particular for the usually dominant
contributions where the additional photon is emitted from the electron line~\cite{Aliberti:2024fpq}.
However, extending the previous
calculations to situations with more external legs becomes
increasingly tedious.

The purpose of this article is to propose a technical procedure that facilitates the combination of the dispersive approach with the use of automated tools for loop calculations. We call this method \emph{disperon QED}. While the proposed
methods are universal, for the presentation of concrete results we
focus on the process \eepp{}. We consider the inclusion of the \ac{VFF}
in loops at NLO and the \ac{HVP} effects restricted to \ac{NNLO} \ac{ISC}, where photons are only exchanged among initial-state electrons.
These effects are implemented in the \mcmule{}
framework~\cite{Banerjee:2020rww,McMule:manual}, a \ac{NNLO} Monte Carlo tool for QED processes that is based on \ac{FKSl} subtraction~\cite{Engel:2019nfw} and \acl{NTS} stabilisation~\cite{Banerjee:2021mty}.
The code and results obtained with it are publicly available
\begin{quote}
    \url{https://mule-tools.gitlab.io/}
\end{quote}

Disperon QED relies on the fact that the dispersive approach transforms a photon propagator into a propagator for a massive vector boson that we dub \emph{disperon}. Its mass
is given by the dispersive parameter and needs to be integrated
over. For moderate values of the mass, the corresponding loop diagrams
will be evaluated with \OpenLoops~\cite{Buccioni:2017yxi,Buccioni:2019sur}. For very large values of 
this mass, we improve numerical stability and speed by turning to
a description inspired by \ac{EFT}
considerations. This custom-made \ac{EFT}, \ac{DET}, is obtained by integrating out the heavy
disperon and keeping terms up to a certain order in the inverse mass. As
we will show, combining these techniques with a threshold subtraction,
we can address the inclusion of \ac{VFF} in the loop in an efficient
manner. In addition, it allows us to not only include a single \ac{HVP}
insertion, but even use a resummed HVP within a loop. While
the effect of resummation is small for the \ac{ISC}
considered here, the resummation can have substantial effect in the
$s$-channel near resonances.

We start in Section~\ref{sec:numinput} with the basic idea of the dispersive
approach and how it relates directly to disperon QED.
In Section~\ref{sec:HVPwdisperon} we address \ac{HVP} insertions, followed in Section~\ref{sec:FsQEDwdisperon}
by \ac{VFF} insertions, to achieve a \ac{FsQED} description. The modifications
that are required in \OpenLoops{} to deal with a disperon are described
in Section~\ref{sec:openloops}. Section~\ref{sec:det} is devoted to the EFT considerations
that are required to deal with the large-disperon-mass region in the
dispersive integral. A final ingredient needed is how to deal with
threshold singularities. This will be done in Section~\ref{sec:threshold}. Then we
are ready to show phenomenological results for the process
\eepp{} in Section~\ref{sec:results}. Apart from presenting our
conclusions in Section~\ref{sec:conclusion}, we also give an outlook for future
applications to more complicated processes. Some technical details are
delegated to several appendices. 

\section{Numerical input functions in loop integrals} \label{sec:numinput}

When calculating higher-order corrections that at any point involve composite particles such as hadrons, we need to solve loop integrals that include functions depending on the loop momentum $k$ but are presumed to be only known numerically.
The way out is to use a dispersive representation of these functions.
Here, we will consider the \ac{HVP} function $\Pih$ or the pion \ac{VFF} $\Fpi$ as examples, though other numerical input functions can be treated the same way. We will use the once-subtracted dispersive representation \eqref{eq:disp} with $s_\text{thr}=\sth$. The interpretation of $k^2=k^2+i\delta$ in \eqref{eq:disp} is always understood, even if often suppressed in the notation. 

We stress that the disperon QED method is not limited to those scenarios but is applicable wherever a dispersion theorem like~\eqref{eq:disp} exists, with a function $F(k^2)$ that fulfils the requirements set out in Appendix~\ref{sec:disp-proof}. For the sake of completeness a proof of~\eqref{eq:disp} is also sketched in Appendix~\ref{sec:disp-proof}.
In the following, we will discuss the use of~\eqref{eq:disp} for (hadronic) \ac{VP} insertions and the calculation of one-loop corrections in \ac{FsQED}.
As an example, we will consider the process $ee\to\pi\pi$ but the same technology can be applied to more complicated processes such as $ee\to\pi\pi + n\gamma$.
Moreover, it enables us to compute the (hadronic) \ac{VP} contribution to any QED process at two loop.

\subsection{HVP with a disperon}\label{sec:HVPwdisperon}
Consider the simple case of a single \ac{VP} insertion. The photon propagator gets modified
\begin{align}
    \frac1{k^2} \to \frac{\Pi(k^2)}{k^2}
    \label{eq:vpsub}
\end{align}
for a suitable routing where the momentum flowing through the \ac{VP} is exactly $k$.
$\Pi$ is renormalised on-shell, i.e.~$\Pi(0)=0$.
In the case of hadrons ($\Pih$) it is related to the measurable $R$-ratio as
\begin{align} \label{eq:Rratio}
    \Im\,\Pih(s) &= -\frac{\alpha}{3} R(s) = -\frac{s}{4\pi\alpha}\sigma(e^+e^-\to {\rm hadrons})\,,
\end{align}
where $\alpha$ is the QED coupling.
Hence, we cannot solve the integral containing~\eqref{eq:vpsub} directly.
Instead, we write, e.g.~\cite{Jegerlehner:2017gek}
\begin{align}
    \frac{\Pih(k^2)}{k^2} = - \frac{1}{\pi} \int_{\sth}^\infty \frac{\D \dpone}{\dpone} \frac{\Im\,\Pih(\dpone)}{k^2-\dpone+\io}\,,
    \label{eq:disphvp}
\end{align}
using \eqref{eq:disp} with $\Pi(0)=0$.
The dispersion integral runs from threshold, i.e.~$s_\text{thr}=\sth$ in the case of \ac{HVP}, to infinity.
The important point is that the $k$ dependence of the r.h.s.~of~\eqref{eq:disphvp} is now of the form of a propagator which allows us to use standard loop calculus.
The new propagator $1/(k^2-\dpone)$ looks like the propagator of a massive photon, a disperon, whose mass $\sqrt{\dpone}$ corresponds to the dispersion parameter.
Since the \ac{HVP} insertion takes place on any photon propagator, we can perform the calculation in a theory that has the disperon as a particle in addition to the photon.
The disperon has the exact same coupling structure as the normal photon and we can use standard loop technology to generate and evaluate diagrams in disperon QED without having to worry that it originated from a dispersive integral.
We note that the disperon has no effect on the underlying gauge symmetry of QED.
Once the loop calculation is completed, we need only to remember to multiply with $\Im\,\Pih(\dpone)$ and accompanying prefactors, and numerically integrate over $\dpone$.
As we will eventually implement this calculation in a Monte Carlo code, we will do this as part of the phase-space integration.

Let us now make the discussion above more concrete.
For the process under consideration, $ee\to\pi\pi$, \ac{VP} contributions can only enter at \ac{NNLO} since the \ac{VP} insertion into a tree-level diagram is already covered by the \ac{VFF}.
The ISC \ac{VP} contribution is given by
\begin{align}
\label{eq:NNLOVPamplitude}
    \mathcal{A}^{(2)}_{\rm VP} = \begin{gathered}
        \begin{tikzpicture}
            \draw (135:1.5) -- (0,0) -- (-135:1.5);
            \draw [photon] (135:1.1) -- (-135:1.1);
            \fill[gray] ({cos(135)*1.1}, 0) circle (0.2);
            \draw [photon] (0,0) -- (1,0);
            \draw [dashed] (1,0) --+ (45:1.5);
            \draw [dashed] (1,0) --+ (-45:1.5);
        \end{tikzpicture}
    \end{gathered}
    \supset
    \int[\D k] \frac{\Pi(k^2)}{\big[k^2\big] \big[(k+p)^2-m_e^2\big] \big[(k-q)^2-m_e^2\big]}\,,
\end{align}
where $[\D k]$ is using the {\tt Package-X}~\cite{Patel:2015tea,Patel:2016fam} convention.
Other contributions, for example when the \ac{VP} is inside a box, are not part of the \ac{ISC} and hence beyond the approximation that is considered within this work.

Before we can use the dispersion relation, let us convince ourselves that the conditions discussed in Appendix~\ref{sec:disp-proof} are fulfilled.
We do this by studying the perturbative case for a particle with mass $m\neq0$ at one loop
\begin{align}
\label{eq:leptonpi}
    \frac{\Pi(k^2)}{k^2} = \frac{\alpha}{18 \pi k^2}\Bigg(6 \beta ^2+3 (\beta ^2-3) \beta  \log\frac{\beta -1+\io}{\beta +1}-16\Bigg)\,,
\end{align}
where we have defined $\beta = \sqrt{1-4m^2/k^2}$.
This function has the necessary branch cut along the real axis from $k^2=4m^2$.
Near threshold ($k^2=4m^2$, $\beta\sim0$), this becomes
\begin{align}
    \frac{\Pi(k^2)}{k^2}\Bigg|_{k^2\sim4m^2} = -\frac{\alpha}{\pi}\frac{2}{9m^2} + \mathcal{O}(\beta)\,,
\end{align}
which is sufficiently well-behaved.
For large $k^2$, we have
\begin{align}
    \frac{\Pi(k^2)}{k^2}\Bigg|_{k^2\sim\infty} = -\frac{\alpha}{9\pi k^2}\bigg(5+3\log\frac{-m^2}{k^2}\bigg)+ \mathcal{O}(k^{-4})\,,
\end{align}
which falls off fast enough for the dispersion relation to be valid.

In the case of $\Pi_h$, the high-energy limit is given by an operator product expansion in perturbative QCD.
The leading contribution, ignoring all $\mathcal{O}{(\alpha_s)}$ terms, is given by the simple vacuum polarisation used in~\eqref{eq:leptonpi}  multiplied with the appropriate prefactors. 
Hence, the limit is well-behaved and reads
\begin{align}
    \frac{\Pi_h(k^2)}{k^2}\Bigg|_{k^2\sim\infty} = \frac{\Pi(k^2)}{k^2} \cdot N_c \sum_f Q_f^2 \Bigg|_{k^2\sim\infty}\,.
\end{align}
At the threshold, $k^2=\sth$, $\Pih(k^2)$ is only accessible through data.
Because the first resonance, the $\rho/\omega$, is not close to the threshold (indeed, it is at $\approx(7-8)\times\,\sth$), we can approximate $\Pih(k^2)$ as a linear function which satisfies the conditions in Appendix~\ref{sec:disp-proof}.

We note that using the simple replacement~\eqref{eq:vpsub} close to narrow resonances such as the $J/\psi$ can be very delicate.
It is often customary to use Dyson resummation instead and write 
\begin{align}
    \frac1{k^2} \to \frac{1}{k^2}\frac{1}{1-\sum_i\Pi_i(k^2)} \equiv \frac{\Pires(k^2)}{k^2}\,,
\end{align}
where the sum is over both leptonic and hadronic \ac{VP}. The resummation can have a large impact in specific regions of observables. A particularly drastic example is given in \cite{Aliberti:2024fpq}, where the invariant mass of the muon pair around the $J/\psi$ region in $ee\to\mu\mu\gamma$ is shown to differ by 60\% between resummed and non-resummed NLO computations. Hence, it is prudent to also include resummation one order higher. This amounts to the computation of box and pentagon diagrams with a Dyson resummed propagator.

Since $\sum_i\Pi_i(k^2)\neq{1}$ for all values of $k^2$, this new $F(k^2)$ has the same analyticity properties and acceptable limiting behaviour
\begin{align}
    \frac{\Pires(k^2)}{k^2}\Bigg|_{k^2\sim4m^2} &= \frac{1}{4 m^2} \frac{1}{1+\frac{8\alpha}{9\pi}} + \mathcal{O}(\beta)\,,\\
    \frac{\Pires(k^2)}{k^2}\Bigg|_{k^2\sim\infty} &= \frac{1}{k^2}\frac1{1+\frac{\alpha}{9\pi}\big(5+3\log\frac{-m^2}{k^2}\big)}+ \mathcal{O}(k^{-4})\,.
\end{align}
We therefore conclude that $\Pires$ is an acceptable $F$ but note that $\Pires(0)=1$ instead of $\Pi(0)=0$ which means that the dispersion relation~\eqref{eq:disphvp} would need to be modified.
To avoid this, we instead define $\Pires'=\Pires-1$ which also avoids mixing up the pure QED vertex correction at \ac{NLO} (the $\Pires(0)=1$ term) and the \ac{NNLO} \ac{VP} effects.
Therefore, we may now use the replacement~\eqref{eq:disphvp} for~\eqref{eq:NNLOVPamplitude} and arrive at
\begin{align}
\label{eq:HVPmidexpressions}
    \mathcal{A}^{(2)}_{\rm VP} \supset
    \frac1\pi \int_{4m_e^2}^\infty \D \dpone\ \frac{\Im\, \Pi(\dpone)}{\dpone}\int[\D k] \frac{1}{\big[k^2-\dpone\big] \big[(k+p)^2-m_e^2\big] \big[(k-q)^2-m_e^2\big]}\,.
\end{align}
Note that in the case of the simple replacement~\eqref{eq:vpsub}, this procedure can also be used for only the hadronic (leptonic) contributions to the \ac{VP}, where the lowest threshold is $\sth$ ($4m_e^2$) instead.
The general form~\eqref{eq:HVPmidexpressions} can be easily written using disperon QED
\begin{align}
    \mathcal{A}^{(2)}_{\rm VP} \supset
    \frac1\pi \int_{4m_e^2}^\infty \D \dpone\frac{\Im \,\Pi(\dpone)}{\dpone}\,\,
    \begin{gathered}
        \begin{tikzpicture}
            \draw (135:1.5) -- (0,0) -- (-135:1.5);
            \draw [zigzag,line width=1.2pt] (135:1.1) -- (-135:1.1);
            \draw [photon] (0,0) -- (1,0);
            \draw [dashed] (1,0) --+ (45:1.5);
            \draw [dashed] (1,0) --+ (-45:1.5);
        \end{tikzpicture}
    \end{gathered}
\end{align}
where we have used a zigzag line to indicate the disperon.

We have thus reduced the inclusion of \ac{VP} contributions to standard amplitudes with a new massive particle. For the computation of such amplitudes, we can use the full machinery available.

\subsection{FsQED with a disperon}\label{sec:FsQEDwdisperon}
The method presented for the inclusion of the \ac{HVP} can be adapted
to any hadronic matrix element depending on a single momentum. However, in more complicated situations we need to include
hadronic matrix elements with more than two external legs. In
particular, for the process \eepp{} at \ac{NLO} the
hadronic matrix element $\gamma^*\,\gamma^*\,\pi\pi$ (with
off-shell photons), i.e.~the (doubly virtual) pion Compton tensor, is
needed. 

Using the full Compton tensor as a building block for the process \eepp{} is trivial, as long as the
tensor does not appear in a loop. If it does, it is currently not
feasible to use the complete expression. The full expression for this matrix element has a kinematic
structure that is extremely difficult to reconcile with standard loop
calculations~\cite{Hoferichter:2025rescatteringpp}. 
Fortunately, as detailed in Appendix~\ref{sec:ComptonTensor}, the
pole part of the pion Compton tensor corresponds to the \ac{sQED}
version, multiplied by the \ac{VFF} $F_\pi^V$ that describes the interaction of an electromagnetic current with a pair of charged on-shell pions and is defined as 
\begin{align}
\langle \pi^+(p) \pi^-(q) \vert j_{em}^\mu(0)\vert 0 \rangle = (-ie) F^V_{\pi}\big((p+q)^2\big)(q^\mu - p^\mu) = 
\begin{gathered}
    \begin{tikzpicture}
        \draw[dashed] (0,0) -- (45:1) node [right] {$\pi^+(p)$};
        \draw[dashed] (0,0) -- (-45:1) node [right] {$\pi^-(q)$};
        \draw[photon] (0,0) -- (180:1) node [left] {$\mu$};
        \fill[gray] (0.15,0.) circle (0.2);
    \end{tikzpicture}
    \end{gathered}\,,
    \label{eq:feynmanrule:fxsqed}
\end{align}
where the blob corresponds to $\Fpi$.
Thus, the use of \ac{FxsQED} for the Compton tensor is a good approximation, at least at low energies.

Once we add a loop integral -- e.g.~by attaching an electron line to the two off-shell photons -- the \ac{FxsQED} Compton tensor leads to a situation akin to the \ac{HVP} case discussed in Section~\ref{sec:HVPwdisperon} where we have the \ac{VFF} in a loop integral.
For the amplitude \eepp{}, using the \ac{FxsQED} Compton tensor leads to a description called \ac{FsQED}.
We generalise this notion and use \ac{FsQED} as a technical definition for the calculation of
any amplitude with external pions, whereby we use \ac{sQED} but
locally (i.e.~also within loops) multiply the photon-pion interaction
with \ac{VFF}. This has to be distinguished from a \ac{FxsQED}
approach, where the \ac{VFF} is multiplied externally, i.e.~after
the loop integration has been carried out. In this latter approach,
the momentum to be used in the \ac{VFF} is ambiguous. Furthermore, as
shown in~\cite{Ignatov:2022iou, Colangelo:2022lzg,Budassi:2024whw}, \ac{FxsQED} does not give a satisfactory description of the forward-backward asymmetry of \eepp{}, that we will define later in~\eqref{eq:CMDasym}.

In~\cite{Aliberti:2024fpq}, FsQED was defined as including the pole terms of the
hadronic matrix elements. For \eepp{}, this agrees with our technical
definition. 
This is also the
case for
\eeppg{} at \ac{NLO} if the additional photon emission is 
from the electron line. However, the correspondence ceases to hold for
\eeppg{} one-loop contributions with additional photon emission from the pion line. As discussed in Appendix~\ref{sec:ComptonTensor}, there is
no one-to-one correspondence between the \ac{FxsQED} version of a
$\pi\,\pi\,n\,\gamma$ matrix element and its pole terms
for $n\ge 3$. Still, in these more complicated cases, the use
of \ac{VFF} within loops is a meaningful first step towards a more
reliable description of processes with external pions~\cite{Kaziukenas:2025gggpp}.  

To summarise, we will
keep using the term \ac{FsQED} for matrix elements evaluated with this
approach and in this section we will describe how to use disperon QED
methods to deal with the corresponding technical issues.  In passing, we also note that using form
factors in loops for $e\,p\to{e}\,p$ does not correspond to keeping
the pole terms in the hadronic matrix element
$\gamma^*\,\gamma^*\,p\,p$. But the notion of \ac{FsQED} as described here
can still be applied and constitutes a starting point for improving
\ac{TPE} contributions in electron-proton scattering.

Turning to the concrete example of \ac{NLO} corrections to $ee\to\pi\pi$, we have \ac{ISC} corrections
\begin{align}
    \mathcal{A}^{(1)}_{\rm ISC} = \begin{gathered}
        \begin{tikzpicture}
            \draw (-1.3,0.8) -- (-0.5,0) -- (-1.3,-0.8);
            \draw[photon] (-0.5,0) -- (0.5,0);
            \draw[photon] (-1.1,0.6) -- (-1.1,-0.6);
            \draw[dashed] (1.3,0.8) -- (0.5,0) -- (1.3,-0.8);
            \fill[gray] (0.5,0.) circle (0.2);
        \end{tikzpicture}
    \end{gathered}\,.
\end{align}
Those corrections are trivial w.r.t.~the inclusion of the \ac{VFF} and do not need to be discussed.

Further, we require so-called mixed contributions with two photons connecting the initial-state electrons with the final-state pions.
Our previous discussion of the Compton tensor allows us to approximate this as
\begin{align}
\label{eq:A1mixed}
     \mathcal{A}^{(1)}_{\rm mixed} = \begin{gathered}
        \begin{tikzpicture}
            \draw (135:1.5) -- (135:1) -- (-135:1) -- (-135:1.5);
            \draw[dashed] (45:1.5) -- (45:1) -- (-45:1) -- (-45:1.5);
            \draw[photon] (135:1) -- (45:1);
            \draw[photon] (-135:1) -- (-45:1);
            \fill[gray] (45:1) circle (0.2);
            \fill[gray] (-45:1) circle (0.2);
        \end{tikzpicture}
    \end{gathered}
    + \begin{gathered}
        \begin{tikzpicture}
            \draw (135:1.5) -- (135:1) -- (-135:1) -- (-135:1.5);
            \draw[dashed] (45:1.5) -- (45:1) -- (-45:1) -- (-45:1.5);
            \draw[photon] (135:1) -- (-45:1);
            \draw[photon] (-135:1) -- (45:1);
            \fill[gray] (45:1) circle (0.2);
            \fill[gray] (-45:1) circle (0.2);
        \end{tikzpicture}
    \end{gathered}
    + \begin{gathered}
        \begin{tikzpicture}
            \draw (135:1.5) -- (135:1) -- (-135:1) -- (-135:1.5);
            \draw[dashed] (45:1.5) -- (0:0.5) -- (-45:1.5);
            \draw[photon] (135:1) -- (0:0.5);
            \draw[photon] (-135:1) -- (0:0.5);
            \fill[gray] (0:0.5) circle (0.2);
        \end{tikzpicture}
    \end{gathered}
    \,,
\end{align}
where we include a \ac{VFF} for each grey blob.
By bringing the $F_\pi^V$ into the loop, we transition from an \ac{FxsQED} description of $\gamma^*\gamma^*\to \pi\pi$ to an \ac{FsQED} description of $ee\to\pi\pi$.
It is possible to fit $F_\pi^V$ to propagators to perform an analytic loop integration as done in the \ac{GVMD} approach.
However, we will not rely on such a parameterisation of $F^V_\pi$.

\Ac{FSC} corrections, i.e.~corrections only associated to the pions,
are not considered within this work.
As discussed in~\cite{Aliberti:2024fpq}, \ac{sQED} may provide a reasonable approximation for those corrections in \eepp{}~\cite{Monnard:2021pvm}.
In \cite{Budassi:2024whw}, these corrections were implemented for \eepp{} using \ac{FsQED} and \ac{GVMD}.
The effects are of a few percent relative to LO for observables that do not require additional photons, e.g.~$\D\sigma/\D\theta$.

The \ac{VFF} vanishes at high energies \cite{Farrar:1979aw}
\begin{align}
\frac{\Fpi(k^2)}{k^2} \Bigg|_{k^2\sim\infty} = \frac{1}{k^2} \frac{8 \pi \alpha_s(k^2) f^2_{\pi} }{k^2}\,,
\end{align}
where $f_{\pi}$ is the pion decay constant and $\alpha_s$ the strong coupling.
This allows us to decompose $F^V_\pi$ dispersively as
\begin{align}
    \frac{\Fpi(k^2)}{k^2} = \frac{1}{k^2} - \frac1\pi\dispint \frac{\D \dpone}{\dpone}\frac{\Im \,\Fpi(\dpone)}{k^2-\dpone}\,.
    \label{eq:dispvff}
\end{align}
In contrast to \eqref{eq:disphvp} we now have $\Fpi(0)=1$.
When we use this to calculate $\mathcal{A}^{(1)}_{\rm mixed}$, we have to consider all combinations of photons and disperons.
In the case of the box diagram, $\mathcal{A}^{(1)}_{\rm mixed}$ is given by
\begin{align}
\label{eq:AfullFsQED}
    \mathcal{A}^{(1)}_{\rm mixed} \supset
    \underbrace{\boxdiag{photon}{photon}}_{pp}
    &- \underbrace{
        \frac1\pi\dispint\D\dpone\frac{\Im \,F_\pi(\dpone)}{\dpone}
        \Bigg(
            \boxdiag{zigzag, line width=1.2pt}{photon} + \boxdiag{photon}{zigzag, line width=1.2pt}
        \Bigg)
    }_{pd}
     \\
    &+\underbrace{
        \frac1{\pi^2}\dispint\D\dpone\D \dptwo
        \frac{\Im \,F_\pi(\dpone)}{\dpone}\frac{\Im \,F_\pi(\dptwo)}{\dptwo}
        \boxdiag{zigzag,line width=1.2pt}{zigzag,line width=1.2pt}
    }_{dd}\,. \nonumber
\end{align}
An analogous split into photon-photon ($pp$), photon-disperon ($pd$), and disperon-disperon ($dd$) contributions is required for the crossed box and the seagull diagram of \eqref{eq:A1mixed}.
In~\eqref{eq:AfullFsQED}, we can now use standard techniques to calculate the box diagrams and obtain expression that can be numerically integrated.
However, there is one subtlety.
In the $pd$ term of the box (this is not the case for the seagull diagram), we have a singularity $(s-\dpone)^{-1}$
\begin{align}
    \mathcal{A}^{(1)\,pd}_{\rm mixed} \sim
        \dispint\D\dpone\frac{\Im \,F_\pi(\dpone)}{\dpone} \Bigg( \underbrace{\boxdiag{zigzag,line width=1.2pt}{photon}}_{\propto (s-\dpone)^{-1}} + \underbrace{\boxdiag{photon}{zigzag,line width=1.2pt}}_{\propto (s-\dpone)^{-1}}\Bigg)\,.
    \label{eq:threshold}
\end{align}
Since $s$ is also above the threshold at $\sth$, we integrate over this singularity.
Such a singularity can arise when the energy of the process, in this case $s$, is just enough to produce the intermediate state on-shell.
Hence, it can only happen in $s$-channel processes.
For $t$-channel processes, the corresponding term would be $\propto(t-\dpone)^{-1}$ and hence unproblematic.
We will come back to the issue of threshold singularities in Section~\ref{sec:threshold}.

\section{Implementation into OpenLoops}
\label{sec:openloops}
A key feature of disperon QED is the possibility to generate and compute the required loop amplitudes with an automated numerical tool, such as \OpenLoops{}~\cite{Buccioni:2017yxi,Buccioni:2019sur}. \OpenLoops{} recursively constructs amplitudes from process-independent building blocks given by the Feynman rules of a model. It provides high numerical stability in one-loop amplitudes due to the on-the-fly reduction of tensor integrals \cite{Buccioni:2017yxi} in combination with a highly efficient numerical stability system \cite{Buccioni:2019sur}. The \OpenLoops{} program provides full QCD and electroweak corrections to the Standard Model, but also allows for the separate calculation of QED and weak corrections with a variable number of massive leptons.

For the applications presented in this paper an extended disperon QED model was implemented in \OpenLoops{}, which contains charged pions, massive leptons, massless photons, and disperons. In addition to the standard QED interactions this includes the \ac{sQED} vertices $\pi\,\pi\,\gamma$ and $\pi\,\pi\,\gamma\,\gamma$, and analogous vertices with disperons.

The applications described in the previous section require the calculation of three types of one-loop amplitudes in this model, namely amplitudes with no disperon, amplitudes with a single disperon of mass $\sqrt{s_1}$ in the loop, and amplitudes with two disperons of independent masses $\sqrt{s_1}$ and $\sqrt{s_2}$ in the loop.
In addition to the kinematics and physical masses, these amplitudes depend on the values $s_i$, which are passed to \OpenLoops{} as numerical parameters.

In order to achieve the splitting of amplitudes into sub-contributions as discussed in Section~\ref{sec:introdution}, we use power counting in the pion and electron charge as well as
diagram selection rules for propagator and vertex types, which are implemented in \OpenLoops{}.

In \OpenLoops{}, divergences are regularised in the 't~Hooft--Veltman scheme~\cite{tHooft:1972tcz}, where external wave functions and momenta are four-dimensional, while loop momenta, metric tensors and Dirac $\gamma$-matrices inside the loops are objects in $ d=4-2\epsilon$ dimensions. 
The numerators of loop integrands are split into a purely four-dimensional part and a $(d-4)$-dimensional remainder. The first is constructed with the \OpenLoops{} algorithm while the latter is reconstructed through the insertion of rational \acp{CT} \cite{Ossola:2008xq,Garzelli:2009is} into tree-level diagrams together with the renormalisation \acp{CT}. 
In the on-shell renormalisation scheme, divergent loop amplitudes with a disperon are renormalised differently from the same amplitudes with a massless photon instead. The renormalisation constants that are required for the electron self-energy, mass and QED vertex are given in Appendix~\ref{sec:renormalisation}. The \ac{UV} pole, however, does not depend on the disperon mass and therefore is the same as in QED. Hence, the rational terms, which depend solely on the \ac{UV} behaviour of the amplitude are also unchanged from QED.

The default regularisation scheme used in \mcmule{} is the four-dimensional helicity scheme~\cite{Bern:2002zk,Kilgore:2012tb,Gnendiger:2014nxa,Gnendiger:2017pys}. Hence, in principle the \OpenLoops{} output has to be adapted to be used within \mcmule. However, for our particular one-loop case, the results in the two schemes coincide. The difference between the 't~Hooft--Veltman scheme and the four-dimensional helicity scheme is induced by collinear singularities. Since we deal with QED with massive matter particles, they are not present.
 
\section{Disperon Effective Theory}
\label{sec:det}

As we discussed above, we can easily use \OpenLoops{} to calculate any required disperon QED amplitude.
However, $\dpone$ in the dispersive integral~\eqref{eq:disp} can go to infinity.
While these contributions are suppressed by $F(\dpone)$, which has to become small for $\dpone\to\infty$, and carry an additional $1/\dpone$ coming from the dispersion relation, we still want a fast and reliable method to evaluate these parts.
For very large values of $\dpone$, the disperon must decouple from the theory which means that the amplitude has to be power suppressed by $\dpone$.
By expanding in this regime, we can guarantee a faster and stable evaluation.
To get the desired level of stability, speed and precision, we need to obtain the $1/\dpone$ (leading power) and $1/\dponesq$ (next-to-leading power) term of the expansion.
This is very similar to the stabilisation technique that \mcmule{} uses to improve the convergence for real-virtual matrix elements.
There we expand up to next-to-leading power in the soft photon energy using \acl{NTS} stabilisation~\cite{Banerjee:2021mty}. 

While it is possible to simply calculate the expanded amplitude using the \ac{MoR}~\cite{Beneke:1997zp}, we instead follow a hybrid approach and combine \ac{MoR} with \ac{EFT}-inspired considerations. We will call this approach \acf{DET}, even though no attempt is made to rigorously construct an EFT. In particular, for our purpose it is not essential to obtain a complete and non-redundant set of higher-dimensional operators.

As it stands, our method will only work easily at one-loop (which means \ac{NNLO} for \ac{HVP} contributions) but it can be extended to higher loops.

The \ac{MoR} means that to all orders in $\dpone$, the amplitude can be decomposed into a hard contribution and a soft contribution.
In an \ac{EFT} language the hard contribution is a tree-level calculation that uses one-loop Wilson coefficients while the soft contribution uses tree-level Wilson coefficients in a one-loop calculation with effective operators.
To perform any disperon QED calculation we are interested in, we need to find dimension-6 (leading power) and dimension-8 (next-to-leading power) operators for all combinations of leptons, photons, and pions.
The construction of a complete basis, e.g.~including the required field redefinitions, beyond dimension 6 is highly non-trivial.
Instead, we propose a hybrid approach where we only use \ac{DET} for the soft contribution while calculating the hard contribution using \ac{MoR}.
This means we will only have to find and match the dimension-6 and dimension-8 operators at tree level.

\subsection{Matching at tree level}

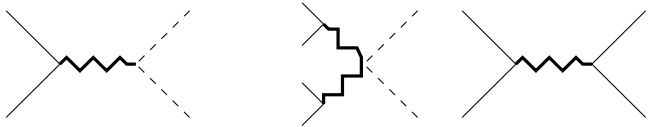
\begin{figure}
    \centering
    \begin{tikzpicture}
        \begin{scope}
            \draw (-1,0) --+ (-135:1);
            \draw (-1,0) --+ ( 135:1);
            \draw[zigzag, line width=1.2pt] (-1,0) -- (0,0);
            \draw[dashed] (-45:1) -- (0,0) --+ (45:1);
        \end{scope}
        \begin{scope}[xshift=3cm]
            \draw (-135:0.75) --+ (-135:0.4);
            \draw (-135:0.75) --+ ( 135:0.4);
            \draw ( 135:0.75) --+ (-135:0.4);
            \draw ( 135:0.75) --+ ( 135:0.4);
            \draw[zigzag, line width=1.2pt] (-135:0.75) -- (0,0) -- (135:0.75);
            \draw[dashed] (-45:1) -- (0,0) --+ (45:1);
        \end{scope}
        \begin{scope}[xshift=6cm]
            \draw (-1,0) --+ (-135:1);
            \draw (-1,0) --+ ( 135:1);
            \draw[zigzag, line width=1.2pt] (-1,0) -- (0,0);
            \draw (-45:1) -- (0,0) --+ (45:1);
        \end{scope}
    \end{tikzpicture}
    \caption{The three tree-level diagrams that need to be calculated for the matching to \ac{DET}.
    The first two diagrams are associated with disperons arising from the \ac{FsQED} approach. In the last one the disperon exclusively interacts with leptons which is due to the \ac{VP} insertion.}
    \label{fig:matching}
\end{figure}

To perform the matching at tree level we need to consider the diagrams shown in Figure~\ref{fig:matching}.
We end up with the following Lagrangian     
\begin{align}
\label{eq:Lmatch}
        \mathcal{L} &\supset
            C^{[6,\pi\pi]}_1\ \bar\Psi\gamma^\mu\Psi \left[ \pi^\dag(-i\overleftarrow{D}_\mu - i \overrightarrow{D}_\mu) \pi \right]
            \\&
          + C^{[8,\pi\pi]}_2\ \left[ \bar\Psi (i\overleftrightarrow{D}_\nu)(i\overleftrightarrow{D}^\nu) \gamma^\mu\Psi \right ]  \left [ \pi^\dag(-i\overleftarrow{D}_\mu - i \overrightarrow{D}_\mu) \pi \right ]
          + C^{[8,\pi\pi]}_3\ (\bar\Psi\gamma^\mu\Psi)(\bar\Psi\gamma_\mu\Psi)\ \pi^\dag\pi \nonumber\\&
          + C^{[6,\Psi\Psi]}_4\ (\bar\Psi\gamma^\mu\Psi)(\bar\Psi\gamma_\mu\Psi)
          + C^{[8,\Psi\Psi]}_5\ \left [ (\bar\Psi\gamma_\mu i\overleftrightarrow{D}_\nu\Psi) \right ] \left [(\bar\Psi\gamma^\mu i\overleftrightarrow{D}^\nu\Psi) \right] \nonumber
          \,,
    \end{align}
where the derivatives only act within the square brackets and $\overrightarrow{D}$ acts on the field to the right. The $\pi$ and $\Psi$ fields contain the annihilation operator of the negatively charged particle. 
The covariant derivative acts as $D_\mu = (\partial_\mu + i e A_\mu)$ on the $\pi$ and $\Psi$ fields.
We furthermore have defined the notation $\pi^\dagger\overleftarrow{D}_\mu\equiv -D_\mu\pi^\dagger$ as well as $\overleftrightarrow{D}_\mu=\overleftarrow{D}_\mu-D_\mu=\overleftarrow{\partial}_\mu-\partial_\mu$.
The particular form of the operator in the first line of~\eqref{eq:Lmatch} was chosen to avoid the necessity of field redefinitions at dimension 8.
The label of the Wilson coefficients indicates the dimension of the corresponding operator and whether it corresponds to the \ac{FsQED} application for disperon QED described in Section~\ref{sec:FsQEDwdisperon} ($\pi\pi$) or it is used to describe the dispersive treatment of \ac{VP} insertions from Section~\ref{sec:HVPwdisperon} ($\Psi\Psi$).
Because we have used gauge covariant derivatives, these operators do not just create $\ell\ell\pi\pi$ or $\ell\ell\ell\ell$ vertices but also $\ell\ell\pi\pi\gamma$.
It is easy to show that the Wilson coefficients take the following values
\begin{align}
    C^{[6,\pi\pi]}_1 &= \frac{4 \pi \alpha }{\dpone}\,,\quad
    C^{[8,\pi\pi]}_2=\frac{ 4 \pi\alpha }{\dpone^2}\,,\quad
    C^{[8,\pi\pi]}_3= \frac{2(4\pi \alpha)^2}{\dpone \dptwo}\,,\quad \nonumber \\
    C^{[6,\Psi\Psi]}_4 &= -\frac{4 \pi \alpha }{\dpone}\,,\quad
    C^{[8,\Psi\Psi]}_5=\frac{4 \pi\alpha }{\dponesq}\,. 
    \label{eq:matchingres}
\end{align}
For the reader's convenience, we include the resulting Feynman rules for the \ac{DET} operators in Appendix~\ref{sec:feynmanrules}.
For all explicit results in this work, the additional factors associated to the dispersion relation for each disperon and their integrals are omitted for simplicity and have to be included in the calculation.

We can now use these Feynman rules to perform one-loop calculations for the \ac{VP} and \ac{FsQED} examples discussed here.
Throughout this work, we rely on {\tt Package-X}~\cite{Patel:2015tea,Patel:2016fam} together with {\tt QGraf}~\cite{Nogueira:1991ex}.
In the case of \ac{FsQED}, we have to distinguish between the case of one heavy and one (or no) light disperon, and the case of two heavy disperons.
As an example, the soft part of the seagull diagram in \ac{FsQED} with one heavy disperon consists of insertions of contributions at dimension 6 as well as dimension 8 from the operators $O^{[6,\pi\pi]}_1$ and $O^{[8,\pi\pi]}_2$ respectively
\begin{align}
    -\int \frac{\D \dpone}{\dpone} \frac{\Im \,\Fpi(\dpone)}{\pi} \seagulls{0.6}\!\implies\!  -\int \frac{\D \dpone}\dpone \frac{\Im \,\Fpi(\dpone)}{\pi} \left( C^{[6,\pi\pi]}_1\seagulleftssix{0.9}+C^{[8,\pi\pi]}_2\seagulleftseight{0.9}\right)\,,
\end{align}
whereas in the case of two heavy disperons only one dimension-8 operator is contributing
\begin{align}
\label{eq:examplesoft}
   \int \int \frac{\D \dpone}\dpone \frac{\D \dptwo}\dptwo \frac{\Im \,\Fpi(\dpone)}{\pi}  \frac{\Im \,\Fpi(\dptwo)}{\pi}
   \seagulld{0.6} \nonumber\\
   \implies \int \int \frac{\D \dpone}\dpone \frac{\D \dptwo}\dptwo \frac{\Im \,\Fpi(\dpone)}{\pi}  \frac{\Im \,\Fpi(\dptwo)}{\pi} \, C^{[8,\pi\pi]}_3 \, \seagulleftdd{0.9}\,.
\end{align}
According to the Feynman rules in Appendix~\ref{sec:feynmanrules}, insertions of an operator $O_j$ are illustrated in Feynman diagrams by a circle around the number $j$.
In the equations above we have illustrated the correct combination of explicit results from~\eqref{eq:matchingres} with the omitted prefactors from the dispersion relation~\eqref{eq:dispvff}.

\subsection{Hard contributions}
Since there are many dimension-8 operators that can be induced at the loop level, we instead
choose to perform the hard calculation in the MoR where all scales except the disperon mass and
the loop momentum are expanded around zero.

In the case of one heavy and one (or no) light disperon, contributions at both dimension 6 and dimension 8 are generated. 
At dimension 6, the result of the box diagram cancels exactly with the result of the crossed box, leaving only a $m^2_e$-suppressed non-zero contribution from the seagull diagram. 
At dimension 8, however, all three diagrams contribute.

In the case of two heavy disperons, applying \ac{MoR} generates no contributions at dimension 6.
Due to the cancellation between box and crossed-box diagrams at dimension 8, the only hard contribution in this case arises from the seagull diagram.
For $e^-(p_1) \, e^+(p_2) \to \pi^- \, \pi^+$, it is given by
\begin{align}
    \seagulld{0.6}
    \,\,\Bigg|^{\rm hard}_{\text{dim8}} 
    =& 
    \frac{\alpha^2 m_e^2 }{\pi^2} \left(\frac{m_e^2 4 (d-1) }{d(2+d) \dponesq (\dpone - \dptwo)} + \frac{s((d-2) \dptwo + (6-d) \dpone)}{2(2+d)\dpone (\dpone - \dptwo)^3}\right)A_0(\sqrt{\dpone})\bar{v}(p_2)u(p_1) \nonumber \\
    &+ (\dpone \leftrightarrow \dptwo)\,.
\end{align}
The resulting tadpole integrals, at one loop given by the Passarino Veltman function $A_0(m)$, can be easily evaluated.

\subsection{Results for DET}

\begin{figure}
    \centering
    \includegraphics[width=0.8\textwidth]{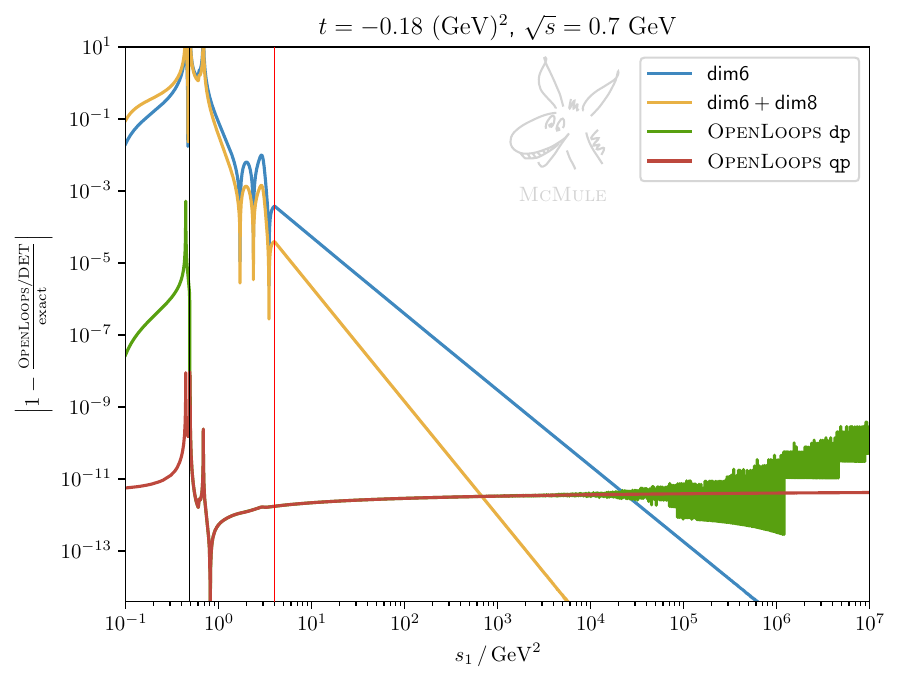}
    \caption{Single-dispersive contribution to the integrand in~\eqref{eq:AfullFsQED} for fixed kinematics but varying the dispersion parameter $s_1$, evaluated using the \ac{DET} at dimension 6 and dimension 8 as well as \OpenLoops{} in \ac{dp} and \ac{qp}.
    The vertical black line represents the value of $\dpone = s$ and hence the location of the threshold singularity.
    The other jumps are due to zero crossings.
    The vertical red line is the value of $s_{\rm cut}$.
    Exact refers to the calculation of the integrand in Mathematica with arbitrary precision.
    }
    \label{fig:single-disp}
\end{figure}

\begin{figure}
    \centering
    \subfloat[
        One of the dispersion parameters, $\dptwo \sim s,t $, is fixed at a small value and indicated with a vertical black line.
        The other parameter is varied.
        ]{\includegraphics[width=0.8\textwidth]{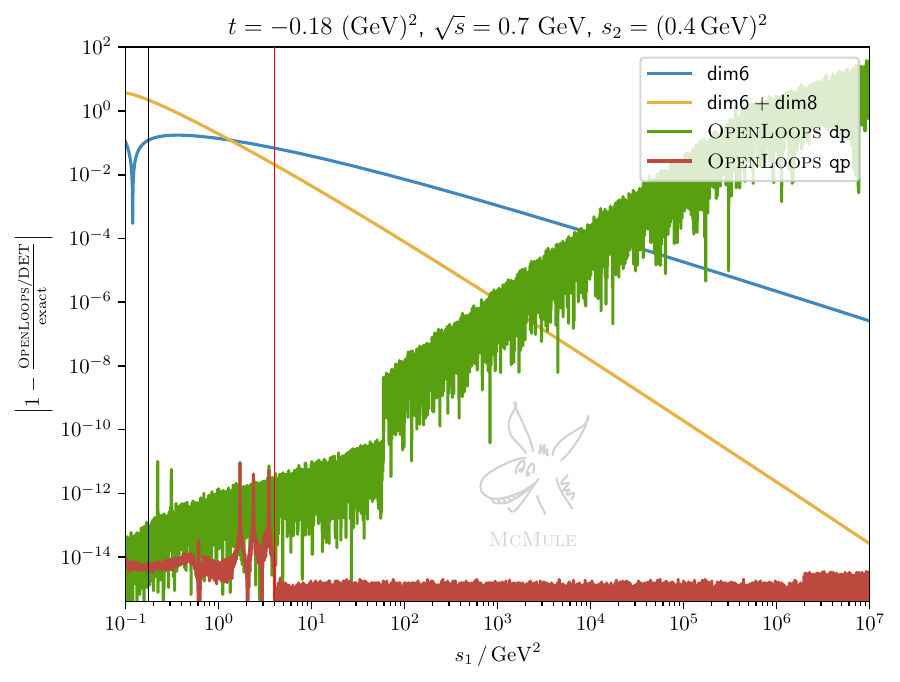}\label{fig:double-single-disp}}\\
    \subfloat[
    Both dispersion parameters are varied with a constant ratio, i.e.~$\dptwo = 2.0 \, \dpone$.
    ]{\includegraphics[width=0.8\textwidth]{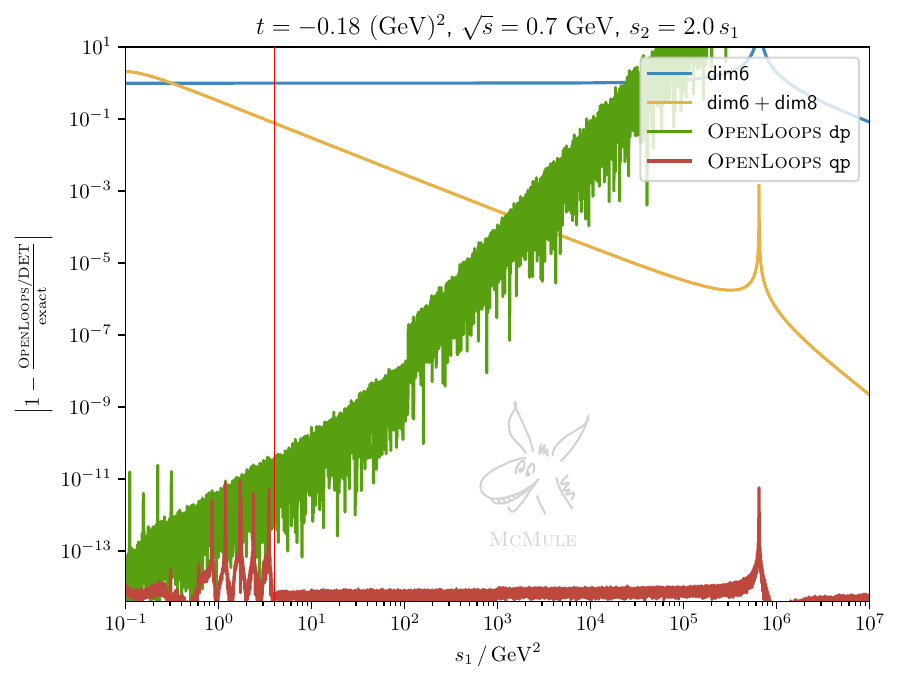}\label{fig:double-double-disp}}
    \caption{
        Double-dispersive contribution for fixed kinematics, similar to Figure~\ref{fig:single-disp}.
    }
    \label{fig:double-disp}
\end{figure}

To validate our method, we compare \ac{DET} at dimension 6 and dimension 8 as well as \OpenLoops{} in \ac{dp} and \ac{qp} against a calculation of the amplitude in Mathematica with arbitrary precision.
To do this, we fix $\sqrt{s}=0.7\,{\rm GeV}$ and $t=-0.18\,{\rm GeV}$ and vary the disperon masses.
Data and plots can be found at~\cite{McMule:data}
\begin{quote}
    \url{https://mule-tools.gitlab.io/user-library/pion-pair/cmd/DETstudy.html}
\end{quote}
Here we consider three scenarios for \eepp{}:
the single-dispersive ($pd$) contribution with a heavy disperon (Figure~\ref{fig:single-disp});
the double-dispersive ($dd$)  contribution with one heavy disperon (Figure~\ref{fig:double-single-disp});
the double-dispersive contribution with two heavy disperons (Figure~\ref{fig:double-double-disp}).
In each scenario, we evaluate the corresponding contribution to the integrand of the dispersive integral in~\eqref{eq:AfullFsQED}.
We first notice that \OpenLoops{} in \ac{qp} is always accurate to at least ten digits which is more than sufficient for our needs.
This also implies that similar tests for \eeppg{} will not require a full calculation in Mathematica for verification.
We also see that \ac{dp} is not always sufficient, especially in the double-dispersive case, for larger values of the dispersion parameter.
Regarding the calculation in \ac{DET}, we notice the power-law scaling in the ratio plots: the slope of the dimension-6 result is much flatter than that of dimension 6 combined with dimension 8 (which includes double insertions of dimension-6 operators).
Furthermore, the double-dispersive contribution with two heavy disperons in Figure~\ref{fig:double-double-disp} clearly shows the necessity of including dimension-8 effects into \ac{DET}.
In that case, the contribution at dimension 6 is exclusively given by the seagull diagrams that are suppressed by $m^2_e$.

Our strategy of using \OpenLoops{} in \ac{dp} below a cut-off value $s_{\rm cut}$ and the \ac{DET} result above this value is therefore validated.
In practice, the value of $s_{\rm cut}$ has to be found for every $\sqrt{s}$ and is chosen such that the final result is independent of it within the numerical error of integration.
An illustration of this procedure can be found in~\cite{Kollatzsch:2026ubi}.
The comparison of \ac{DET} and \OpenLoops{} as in Figures~\ref{fig:single-disp} and~\ref{fig:double-disp} provides a first estimate for $s_{\rm cut}$.
However, several Monte Carlo runs with different $s_{\rm cut}$ are required to confirm the independence of the final result.
For $\sqrt{s}=0.7\,{\rm GeV}$, we find $s_{\rm cut}=4\,{\rm GeV}^2$ to be sufficient for Monte Carlo usage.

Finally, we should point out that for large values of the dispersion parameters, the additional suppression of the integrand~\eqref{eq:dispvff} with $1/\dpone$ as well as the \ac{VFF} suppression start to matter since $\Im\,\Fpi(\dpone) \to 0$ as $\dpone\to\infty$.
Therefore, we can tolerate larger errors in the regime of large $\dpone$.
Using the \ac{DET} result is beneficial as it is much faster than \OpenLoops{} (roughly one order of magnitude compared to \ac{dp} and two orders of magnitude compared to \ac{qp}).

 
\section{Threshold singularities}
\label{sec:threshold}

We now turn our attention to the threshold singularity we first encountered in~\eqref{eq:threshold}.
The problem is that the disperon-photon box in the $pd$ contribution for \eepp{} has a $(s-\dpone)^{-1}$ singularity.
As we are numerically integrating over $\dpone$ from the threshold $\sth<s$, we will hit this singularity.
This is not a new problem.
Indeed, it was already noted when calculating the \ac{HVP} corrections to $s$-channel Bhabha scattering using what we would now call disperon QED~\cite{Kuhn:2008zs}.
The underlying box diagram is the same, except that it gets multiplied with $R(\dpone)$ from~\eqref{eq:Rratio} rather than $\Im\,\Fpi(\dpone)$.
In \cite{Kuhn:2008zs}, the solution was to set $R(\dpone) \to R(\dpone)-R(s)$ in the numerical integration and correct for this analytically, i.e.~they write
\begin{align}
    \dispint\D\dpone\frac{R(\dpone)}{\dpone} \disperonphotonbox =
    \dispint\D\dpone\frac{R(\dpone)-R(s)}{\dpone} \disperonphotonbox + \dispint\D\dpone\frac{R(s)}{\dpone} \disperonphotonbox\,,
    \label{eq:kuhnsubtraction}
\end{align}
where the second integral on the r.h.s~is performed analytically with a regulator.
Since for physical applications only the real part of the product of~\eqref{eq:kuhnsubtraction} and the tree-level amplitude enters, \eqref{eq:kuhnsubtraction} works well if the loop integrals are the only source of imaginary parts.
However, in the case of \eepp{} where the complex \ac{VFF} of the Born amplitude contributes further imaginary parts, also the imaginary part of the last integral in~\eqref{eq:kuhnsubtraction} has to be evaluated.
For the real part, this integral is well defined as the behaviour above and below the $s=\dpone$ threshold cancels.
Since the box diagram becomes imaginary only for $\dpone < s$, the principal-value prescription no longer cancels the singularity, and a different method must be used.

In~\cite{Colangelo:2022lzg,Budassi:2024whw} the real and imaginary parts of $\Fpi(\dpone)$ are treated separately.
The two groups both use a subtraction similar to \cite{Kuhn:2008zs} and then either use the analytic regulator for the endpoint divergence~\cite{Colangelo:2022lzg} 
or Cutkosky rules~\cite{Cutkosky:1960sp} to calculate the imaginary part~\cite{Budassi:2024whw}.

Here, we opt for an entirely different approach which promises greater universality by constructing a \ac{CT} for the whole amplitude that can be generalised to more complicated topologies.

We will first present our approach in Section~\ref{sec:Landauanalysis} and calculate the explicit \ac{CT} for \eepp{} in Section~\ref{eq:constrctCT}.
Afterwards, we will discuss its interplay with \ac{IR} divergences in Section~\ref{sec:IRdiv} and conclude in Section~\ref{sec:universalCT} that an adaption of the CT to more complicated processes is straightforward.

\subsection{Landau analysis}\label{sec:Landauanalysis}

To systematise our analysis, we perform a Landau analysis to determine the location of the kinematic singularity on the relevant integral.
For \eepp{}, this is
\begin{align}
    I = \boxdiag{zigzag, line width=1.2pt}{photon} =  \int \frac{[\D k]}{\big[(k-p_1-p_2)^2-s_1\big] \big[k^2-2 k\cdot p_2\big] \big[k^2\big] \big[k^2-2 k\cdot p_4\big]}\,,
    \label{eq:int:ee2uu}
\end{align}
with $p_1^2=p_2^2 = m_e^2$, $(p_1+p_2)^2=s$, $(p_2-p_4)^2=t$ and $p_4^2=m_\pi^2$.
Employing the Landau equations~\cite{Landau:1959fi,Eden:1966dnq} for an $N$-propagator integral
\begin{align}
    \label{eq:landaueq}
    x_i \frac{\partial\F}{\partial x_i} = 0\qquad\forall i \in \{ 1, ... N \} \quad \text{without all } x_i=0\,,
\end{align}
where the $\mathcal{F}$ graph polynomial of~\eqref{eq:int:ee2uu} is given by
\begin{align}
    \F =
        m_e^2 x_1 (x_1+x_4)+m_\pi^2 x_4 (x_1+x_4)+(-t) x_1 x_4
        +\dpone x_2 (x_1+x_2+x_3+x_4)
        -s x_2 x_3\,,
\end{align}
does not disclose a singularity.
In general, a more careful analysis is necessary to find all singularities of Feynman integrals, e.g.~\cite{Fevola:2023kaw}.
To that end, we employ the {\tt PLD.jl} package~\cite{Fevola:2023kaw,Fevola:2023fzn} together with {\tt PLD-Wrapper}~\cite{PLDwrapper}.
This tool enables searches for singularities by allowing for a scaling
\begin{align}
\label{eq:scaling}
    x_i \to \varepsilon^{w_i} x_i \quad \text{with} \quad \varepsilon \to 0\,,
\end{align}
instead of just $x_i = 0$ or $x_i \neq 0$ as in~\eqref{eq:landaueq}.
For the integral in~\eqref{eq:int:ee2uu} we find
\begin{align}
    \vec w = (-1, -1, -1, -3) \quad \text{at} \quad \dpone = s\,.
\end{align}
Having all $w_i \neq 0$ means that all propagators pinch simultaneously according to~\eqref{eq:scaling}.
Hence, $\dpone = s$ is a singularity of the whole diagram with no contracted propagators (corresponding to setting the corresponding $x_i = 0$).

Let us now study the integral near the singular point using the \ac{MoR}, i.e.~we write $\dpone=s(1-\lambda)$ and expand for $\lambda\to0$.
By utilising automatised tools like {\tt asy}~\cite{Pak:2010pt,Jantzen:2012mw} or {\tt pySecDec}~\cite{Heinrich:2021dbf}, we find two region vectors
\begin{align}
    \vec v_h = (0,0,0,0)
    \quad\text{and}\quad
    \vec v_s = (-1,-1,-2,-1)\,.
\end{align}
These are defined similarly to the $w_i$ of~\eqref{eq:scaling} as $x_i\to \lambda^{v_i} x_i$ and quantify how each propagator will scale in the small parameter $\lambda$.
This Feynman parameter space representation can be related to momentum space.
It is easy to see that $\vec v_s$ describes a soft region (with scaling $k\to \lambda \, k$) that starts at $\lambda^{-1-2\epsilon}$ while the hard region starts at $\lambda^{0}$.
Since the logarithmically divergent terms are integrable, 
the \ac{CT} only needs to capture the $\lambda^{-1-2\epsilon}$ singular behaviour, i.e.~the soft term.
In momentum space, this can be written as
\begin{align}
    I_s &= \frac{1}{\lambda^{1+2\epsilon}}\int\frac{[\D k]}{\big[2k\cdot(p_1+p_2) + s\big] \big[2k\cdot p_2\big] \big[k^2\big] \big[2k\cdot p_4\big]}\equiv \frac1{\dpone}\Big(\frac{s}{\dpone-s-\io}\Big)^{1+2\epsilon}f(s,t,m_e,m_\pi)\,.
    \label{eq:int:ee2uu:ex}
\end{align}
The \ac{MoR} already tells us the dependence on $\dpone-s$ which we have included above.
The sign of $i\delta$ is due to its appearance in the original integral~\eqref{eq:int:ee2uu} as $(\dots -\dpone + i \delta)$.
The additional $1/\dpone$ has been factored out to ensure the convergence of the dispersive integral.
The remaining factor $f(s,t,m_e,m_\pi)$ still needs to be calculated explicitly to construct a \ac{CT}.
While it is possible to calculate this integral directly, it is easier to simply expand the analytic expression for the full box function.
We find
\begin{align}
    \label{eq:fexplicit}
            f = &  \Big(\frac{\mu^2}{s}\Big)^\epsilon \frac{\chi_t }{\chi_t^2 - 1} \Bigg\{
                \frac{\log (-\chi_t)}{m_e m_\pi }  \Big(\frac{1}{\epsilon }-2 \log (1-\chi_t^2)\Big)
                 \\&
                + \frac{(1+\chi_e^2)(1+\chi^2_\pi)}{s \, \chi_e \chi_\pi } \bigg[ \zeta_2 - \log (\chi_e)^2 - \log (\chi_\pi)^2- \text{Li}_2\Big(\chi^2_t\Big)
                 \nonumber \\&
                -\text{Li}_2\Big(1+\chi_e \chi_\pi \chi_t\Big)
                -\text{Li}_2\Big(1+\frac{\chi_t \chi_\pi}{\chi_e}  \Big)
                -\text{Li}_2\Big(1+\frac{\chi_t}{\chi_e \chi_\pi}  \Big)
                -\text{Li}_2\Big(1+\frac{\chi_t \chi_e}{\chi_\pi}  \Big) \bigg]
            \Bigg\}\,, \nonumber
    \end{align}
    where we have introduced the following symbols
    \begin{align}
        \chi_e^2 = \frac{\sqrt{s}-\sqrt{s-4m_e^2}}{\sqrt{s}+\sqrt{s-4m_e^2}}\,, \quad 
        \chi_\pi^2 = \frac{\sqrt{s}-\sqrt{s-4m_\pi^2}}{\sqrt{s}+\sqrt{s-4m_\pi^2}}\,,\quad 
        \chi_t = \frac{1-\sqrt{\frac{(m_e+m_\pi)^2-t}{(m_e-m_\pi)^2-t}}}{1+\sqrt{\frac{(m_e+m_\pi)^2-t}{(m_e-m_\pi)^2-t}}}\,.
    \end{align}
The same analysis shows that all other integrals related to this diagram are at most logarithmically divergent.

\subsection{Constructing the counterterm}\label{eq:constrctCT}
We are now ready to construct the \ac{CT} for the threshold singularity.
Instead of subtracting the numeric function $F(\dpone)$ as in~\eqref{eq:kuhnsubtraction}, we subtract the whole amplitude, i.e.~we write
\begin{align}
    \dispint\D\dpone\frac{F(\dpone)}{\dpone} \disperonphotonbox = \dispint\D\dpone\Bigg(\frac{F(\dpone)}{\dpone}\disperonphotonbox-\frac{F(s)}{s} \, {\rm CT}\Bigg) + \frac{F(s)}{s}\dispint\D\dpone\ {\rm CT}\,,
    \label{eq:oursubtraction}
\end{align}
The \ac{CT} is now simply given by the singular part of the box at $\dpone=s$
\begin{align}
    {\rm CT} = \eqref{eq:int:ee2uu:ex} = \boxdiag{zigzag, line width=1.2pt}{photon} +  \mathcal{O}(\dpone-s)^0\,.
\end{align}
With this, we have for each single-dispersive term in~\eqref{eq:AfullFsQED} a subtraction of the following form
\begin{subequations}
\begin{align}
    \mathcal{A}^{(1)\,pd}_{\rm mixed} &\supset -\frac{1}{\pi}\dispint\D\dpone\Bigg(
        \frac{\Im \,F_\pi(\dpone)}{\dpone} \disperonphotonbox
        - \frac{\Im\, F_\pi(s)}{s} \frac{1}{\dpone}\Big(\frac{s}{\dpone-s-\io}\Big)^{1+2\epsilon} f(s,t,m_e,m_\pi)
    \Bigg)\label{eq:counterterm1}\\&\qquad
    - \frac{1}{\pi} \frac{\Im \,F_\pi(s)}{s} f(s,t,m_e,m_\pi) \dispint\frac{\D\dpone}{\dpone}\Big(\frac{s}{\dpone-s-\io}\Big)^{1+2\epsilon}\,.
    \label{eq:counterterm2}
\end{align}
\end{subequations}
While~\eqref{eq:counterterm1} will be integrated numerically within the Monte Carlo, the integration of~\eqref{eq:counterterm2} requires our attention.
Note that this integration does not depend on $f$.
We perform a substitution $\dpone \to \sth(1+\tau)$ which casts the integral into the form
\begin{align}
    \dispint\frac{\D\dpone}{\dpone}\Big(\frac{s}{\dpone-s-\io}\Big)^{1+2\epsilon}
    = (1-x)^{1+2\epsilon} \int_0^\infty\D \tau\frac{(x+\tau-\io)^{-1-2\epsilon}}{1+\tau}
\end{align}
with $x=1-s/\sth$.
This integral is a well-known hypergeometric function 
\begin{align}
    \pFq{2}{1}{a,b}{c}{1-\alpha} = \frac{\Gamma(c)}{\Gamma(b)\Gamma(c-b)} \int_0^\infty\D \tau \, \tau^{-b+c-1}\ (1+\tau)^{a-c}\ (\tau+\alpha)^{-a} \quad \text{with} \quad \Re \, c > \Re\, b > 0\,.
\end{align}
Using {\tt HypExp}~\cite{Huber:2005yg,Huber:2007dx} we arrive at
\begin{subequations}
\label{eq:intCT}
\begin{align}
    \dispint\frac{\D\dpone}{\dpone}\Big(\frac{s}{\dpone-s}\Big)^{1+2\epsilon}
    &= \frac{\Gamma(1+2\epsilon) }{\Gamma(2+2\epsilon)}(1-x)^{1+2\epsilon}  \pFq{2}{1}{1+2\epsilon,1+2\epsilon}{2+2\epsilon}{1-x+\io}\\
    \label{eq:intCTepsexpanded}
    &= (1-x)^{2\epsilon} \Bigg[
        \log \frac{1}{x-\io}
        +\epsilon\Big(
            \log^2\frac{1}{x-\io} - 2\text{Li}_2(1-x+\io)
        \Big)
        +\mathcal{O}(\epsilon^2)
    \Bigg]\,.
\end{align}
\end{subequations}
Note that the $\mathcal{O}(\epsilon)$ part of~\eqref{eq:intCT} is required as it will meet the $1/\epsilon$ (\ac{IR}) pole of $f$.
It is now trivial to assemble the integrated \ac{CT} and implement our threshold treatment~\eqref{eq:oursubtraction} in a Monte Carlo.

\subsection{Interplay with IR divergences}\label{sec:IRdiv}
\def\ieik{\hat{\mathcal{E}}}
\newcommand{\Mmix}[1]{\mathcal{M}_{\rm mixed}^{(1)#1}}
\newcommand{\Mborn}{\mathcal{M}^{(0)}}
To obtain a physical cross section, the \ac{IR} poles of the virtual amplitude need to cancel against those of the real phase space integration.
In other words, the \ac{IR} structure of $\Mmix{}=2\,\Re(\mathcal{A}_{\rm mixed}^{(1)}\cdot \mathcal{A}^{(0)*})$ needs to match what is predicted by the Yennie-Frautschi-Suura~\cite{Yennie:1961ad} theorem.

In \mcmule{} the \ac{IR} structure is handled using \ac{FKSl} subtraction and we will use the same notation as in~\cite{Engel:2019nfw} for the \ac{IR} pole here.
To be precise, the pole in dimensional regularisation is
\begin{align}
    \label{eq:YFSrelation}
    \Mmix{}\Big|_\text{IR div.} \stackrel!= -\frac\alpha\pi\frac{1}{2\epsilon}\bigg(
        \frac1{\beta_u}\log\frac{1+\beta_u}{1-\beta_u}
        +\frac1{\beta_t} \log\frac{1+\beta_t}{1-\beta_t}
    \bigg)\Mborn
    \equiv
    -\ieik\Mborn\,,
\end{align}
with the tree-level result $\Mborn$ and $\beta_{Q^2}=\sqrt{1-4m_e^2m_\pi^2 / (m_e^2 + m_\pi^2 - Q^2)^2}$.
We will now discuss all \ac{IR}-divergent contributions ($pp$ and $pd$; $dd$ is finite) and how they interact with each other to demonstrate that~\eqref{eq:YFSrelation} is still valid in \ac{FsQED}.
A similar discussion using a photon mass regulator can be found in~\cite{Colangelo:2022lzg,Budassi:2024whw} where certain sub-leading effects in the regulator need to be considered.
This is not necessary in our scheme. 

The simplest \ac{IR}-divergent term is the non-dispersive part
\begin{align}
\label{eq:IRdivsqed}
    \Mmix{\, pp} \Big|_{{\text{IR div.}}} = -  \frac{\ieik\Mborn}{F_\pi(s)} \,.
\end{align}
It consists of the expected kinematical terms.
One \ac{VFF} of the tree-level result has to be divided out since it is not present in the $pp$ amplitude originating from the dispersion relation~\eqref{eq:dispvff}.
The structure of the divergence of~\eqref{eq:counterterm2} is given by the $\mathcal{O}(\epsilon^0)$ part of the integrated \ac{CT}~\eqref{eq:intCTepsexpanded}. We find
\begin{align}
\label{eq:IRdivpdsecond}
        \big(\eqref{eq:counterterm2} \subset \Mmix{\,pd}\big)\Big|_{{\text{IR div.}}}  & = \left( - \frac{\ieik\Mborn}{F_\pi(s)} \right) \frac{\Im\, F_\pi(s)}{\pi} \log \frac{1}{x} \,.
\end{align}
In addition to the logarithm, we obtain a form factor mismatch between the tree-level result and the $pd$ amplitude.
The most intricate part is the combination of the $pd$ amplitude and the threshold \ac{CT}. 
Thanks to the subtractive structure of~\eqref{eq:oursubtraction}, we arrive at
\begin{subequations}
\label{eq:IRdivpdfirst}
\begin{align}
\label{eq:IRdivfirstintegrandform}
     \left(\eqref{eq:counterterm1}\subset \Mmix{\,pd} \right)\Big|_{{\text{IR div.}}} 
    &= \left( - \frac{\ieik\Mborn}{F_\pi(s)} \right) \frac{s}{\pi}\dispint\frac{\D\dpone}{\dpone} \frac{\Im \,F_\pi(s)-\Im \,F_\pi(\dpone)}{(s-\dpone)}  \\
    &= \left( - \frac{\ieik\Mborn}{F_\pi(s)} \right) \left( - 1 + F_\pi(s) - \frac{\Im \,F_\pi(s)}{\pi} \log \frac{1}{x} \right)\,. 
\end{align}
\end{subequations}
The integrand in~\eqref{eq:IRdivfirstintegrandform} is a sum of two terms. 
The one involving $\Im \, F_\pi(\dpone)$ is the l.h.s.~of the dispersion relation~\eqref{eq:dispvff} and hence leads to $\propto F_\pi(s)$.
The other term, once integrated according to~\eqref{eq:intCT}, will generate the logarithm. 
In the end, we find
\begin{align}
   \Mmix{}\Big|_{{\text{IR div.}}} =  \Mmix{\,pp}\Big|_{{\text{IR div.}}} + \Mmix{\,pd} \Big|_{{\text{IR div.}}} = \eqref{eq:IRdivsqed} + \eqref{eq:IRdivpdsecond} + \eqref{eq:IRdivpdfirst} = -\ieik\Mborn\,.
\end{align}
Hence, we have established~\eqref{eq:YFSrelation} as required.
The amplitude containing the corresponding real corrections does not require the disperon QED framework; it is trivial w.r.t.~the treatment of the \ac{VFF}.

\subsection{Generalising beyond \texorpdfstring{\eepp{}}{ee->pipi}}\label{sec:universalCT}
\begin{figure}
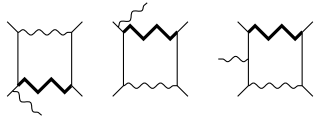

    \centering
    \boxdiagphoton{photon}{zigzag, line width=1.2pt}
    \boxdiagphotontwo{zigzag, line width=1.2pt}{photon}
    \pentagon{zigzag, line width=1.2pt}{photon}
    \caption{Topologies contributing to the threshold singularity in \eeppg{}.}
    \label{fig:ee2uug}
\end{figure}

When considering similar topologies for $ee\to\pi\pi\gamma$ as shown in Figure~\ref{fig:ee2uug}, we find that the singularity is at $\dpone=(p_1+p_2-p_5)^2$.
Singularities arising from box topologies (the first two of Figure~\ref{fig:ee2uug}) can easily be dealt with through the \ac{CT} for \eepp{}.
When analysing the scaling vector $w_i$ of the pentagon topology (last topology of Figure~\ref{fig:ee2uug})
\begin{align}
    I = \int \frac{[\D k]}{
         \big[(k+p_1+p_2-p_5)^2-\dpone\big]
         \big[k^2+2k\cdot(p_2-p_5)-2p_2\cdot p_5\big]
         \big[k^2+2k\cdot p_4\big]
         \big[k^2\big]
         \big[k^2+2k\cdot p_2\big]
     }\,,
\end{align} 
we find that it has one zero entry.
This indicates that the singularity arises from a subdiagram.
In order to show this, we find the soft region vector for the first non-zero power in $\lambda$ 
\begin{align}
     \vec v_s = (-1,0,-1,-2,-1)\,,
 \end{align}
and expand the integral accordingly.
We find
\begin{align}
    I_s = \int \frac{[\D k]}{
        \big[ 2k\cdot(p_1+p_2-p_5) + (p_1+p_2-p_5)^2 \big]
        \big[ -2p_2\cdot p_5 \big]
        \big[ 2k \cdot p_4\big]
        \big[k^2\big]
        \big[ 2k \cdot p_2\big]
    }\,.
\end{align}
Note that the second propagator now no longer depends on $k$ and can therefore be factored out.
Therefore, this integral can be mapped onto the previously calculated \ac{CT}~\eqref{eq:int:ee2uu:ex}.
This means that, apart from momentum shifts in~\eqref{eq:int:ee2uu:ex}, the \ac{CT} also remains unchanged for the pentagon topology.
A similar result can be obtained by first writing the pentagon integral in terms of boxes using Cayley reduction~\cite{Denner:2002ii,Bern:1992em,Bern:1993kr} in which case the mapping to the box integral~\eqref{eq:int:ee2uu} becomes easy to see.

The analysis can be repeated for $ee\to\pi\pi\gamma\gamma$ and finds the same result: the \ac{CT} that was constructed above is universal.
It can be used to treat threshold singularities in all processes (potentially) covered by the disperon QED framework, such as $\tau\to\pi\pi\nu_\tau$ described in FsQED~\cite{Colangelo:2025iad,Colangelo:2025ivq}.

\section{Phenomenological results}\label{sec:results}
Additionally to the technical results presented in the previous sections, we will now present some phenomenological results for \eepp{} that validate our method.
We use the following notation to denote a cross section calculated at N$^n$LO
\begin{align}
    \sigma_n = \sigma^{(0)} + \sigma^{(1)} + \dots + \sigma^{(n)}\,.
\end{align}
In \mcmule{}, the following contributions are available
\begin{subequations}
    \begin{align}
    \sigma^{(1)} &\supset \sigma_{\rm ISC}^{(1)} + \sigma_{\rm mixed\,FsQED}^{(1)}\,, \\
    \sigma^{(2)} &\supset \sigma_{\rm ISC}^{(2)} \supset \sigma_{\rm ISC\,VP}^{(2)}\,,
\end{align}
\end{subequations}
where the \ac{VP} contribution at \ac{NNLO}, $\sigma_{\rm ISC\,VP}^{(2)}$, contain only a single \ac{VP} insertion.
We stress that some of these results have been presented before and that most of them are not novel.
Specifically, the mixed \ac{NLO} corrections have been calculated in \ac{FsQED} in~\cite{Colangelo:2022lzg,Budassi:2024whw}, using \ac{GVMD} in~\cite{Ignatov:2022iou,Budassi:2024whw,Price:2025fiu} and using F$\times$sQED in~\cite{Hoefer:2001mx,Arbuzov:2020foj,Budassi:2024whw}.
We have compared our result for the virtual corrections to \cite{Colangelo:2022lzg}.
The electronic \ac{NNLO} corrections have mostly been presented in~\cite{Aliberti:2024fpq} using \mcmule{}'s calculation of electronic corrections for $e\mu\to e\mu$~\cite{Banerjee:2020rww}.
What is new here is that we are now able to resum the \ac{VP} insertion as discussed above.

Specifically, we will consider the process \eepp{} in the CMD-like scenario of~\cite{Aliberti:2024fpq}, i.e.~with $\sqrt{s}=0.7\,{\rm GeV}$.
We further define
\begin{align}
\label{eq:thav}
    1\,{\rm rad} &\le \thavg = \frac{(\theta^- - \theta^+ + \pi)}{2} \le \pi-1\,{\rm rad}\,,
\end{align}
where $\theta^\pm$ is the angle of the $\pi^\pm$.
In addition we require
\begin{align}
    \begin{split}
        |\vec p_\pm| &> 0.45\times\sqrt{s}/2\,,\\
        \delta\phi &= \big||\phi^+-\phi^-|-\pi\big| < 0.15\,{\rm rad}\,,\\
        \xi &= \big| \theta^++\theta^--\pi\big| < 0.25\,{\rm rad}\,.
    \end{split}
\end{align}
Here we have also used the azimuthal angle $\phi^\pm$ and three-momenta $\vec p_\pm$ of the $\pi^\pm$.
The pion \ac{VFF} is taken from~\cite{Colangelo:2018mtw}, however we stress that \mcmule{} can deal with any \ac{VFF} that is suitable for \ac{FsQED}.

\begin{figure}
    \centering
    \includegraphics[width=\linewidth]{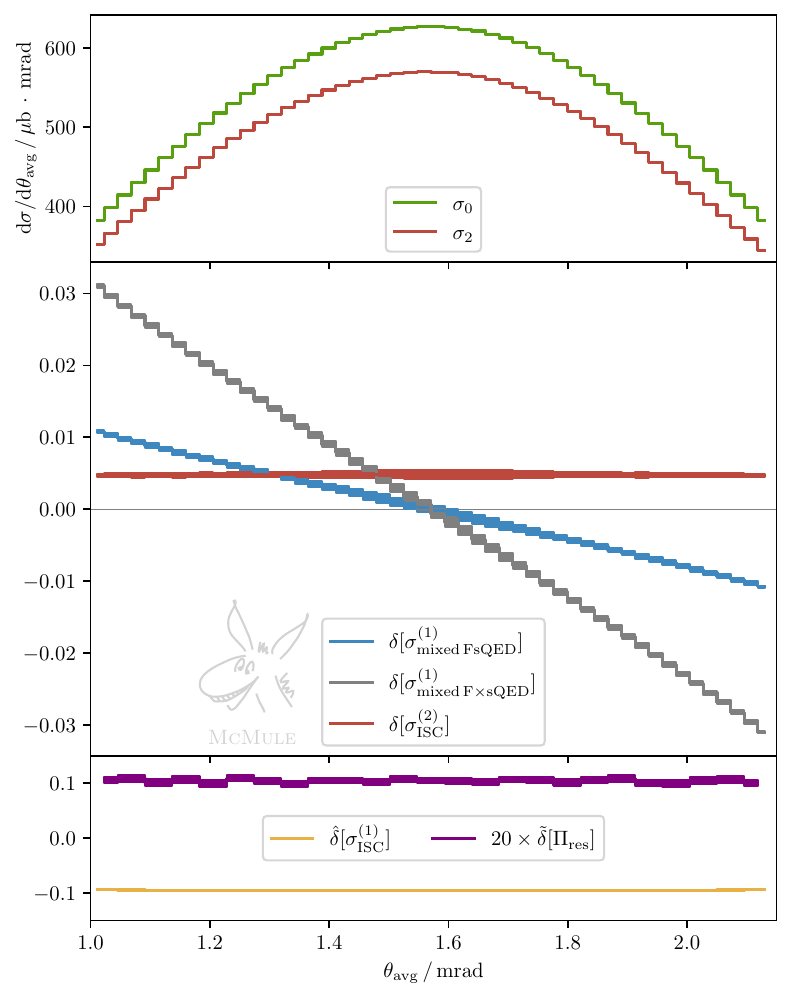}
    \caption{
    The top panel shows the distribution for $\thavg$ defined in~\eqref{eq:thav} at \ac{LO} and the best prediction within \mcmule{}.
    The other panels show the various $\delta$'s defined before.
    Note that the purple line corresponds to a \ac{NNNLO} effect that results in a $10^{-6}$ effect relative to \ac{LO}.
    }
    \label{fig:thav}
\end{figure}

\begin{figure}
    \centering
    \includegraphics[width=\linewidth]{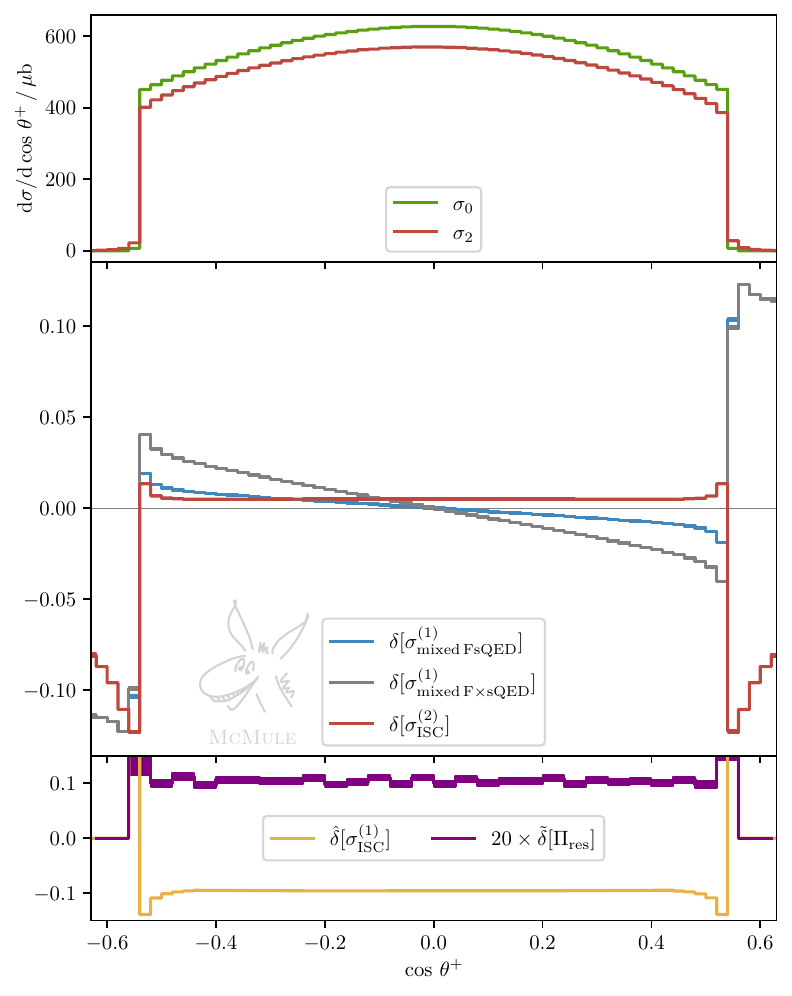}
    \caption{Same plot as in Figure~\ref{fig:thav} but for $\cos\theta^+$.}
    \label{fig:costh}
\end{figure}

By studying the impact of mixed NLO corrections ($\sigma^{\rm mixed}_1 = \sigma^{(0)}+\sigma^{(1)}_{\rm ISC} + \sigma_{\rm mixed\,FsQED}^{(1)}$) in \ac{FsQED}, as well as the effect of resummation of the \ac{VP}, we illustrate the size of contributions that are covered by the disperon QED method.
Due to the number of photons coupling to pions, we note that the mixed corrections only contribute to $C$-odd observables such as $\theta^{\pm}$ and functions thereof. 
As a result, we only show such observables in this work.
All raw data and analysis pipelines can be found at~\cite{McMule:data}
\begin{quote}
    \url{https://mule-tools.gitlab.io/user-library/pion-pair/cmd/analysis.html}
\end{quote}
Figures~\ref{fig:thav} and~\ref{fig:costh} show the size of different contributions w.r.t.~\ac{NLO} \ac{ISC} ($\sigma^{\rm ISC}_1 = \sigma^{(0)}+\sigma^{(1)}_{\rm ISC}$) defined by
\begin{align}
    \label{eq:thedelta}
    \delta[X] = \frac{\sigma^{\rm ISC}_1 + X}{\sigma^{\rm ISC}_1} - 1\,,
\end{align}
where $X\in\{\sigma^{(1)}_{\rm mixed\,FsQED},\,\sigma^{(1)}_{\rm mixed\,F{\times}QED},\,\sigma^{(2)}_{\rm ISC}\}$.
We show the impact of $\sigma^{(1)}_{\rm mixed\,F{\times}QED}$ just for illustration.
For completeness, in the bottom panel we also show the size of $\sigma^{\rm ISC}_1$ compared to $\sigma_0$
\begin{align}
   \label{eq:NLOISCimpact}
    \quad \hat{\delta}[{\sigma^{(1)}_{\rm ISC}}] = \frac{\sigma^{\rm ISC}_1}{\sigma_0} - 1\,,
\end{align}
as well as the impact on the NNLO prediction of using a resummed \ac{VP} compared to a single \ac{VP} insertion
\begin{align}
\label{eq:resumdelta}
    \tilde{\delta}[\Pires] = \frac{\sigma_{\rm ISC\,resummed\,VP}^{(2)}}{\sigma_{\rm ISC\,VP}^{(2)}}-1\,.
\end{align}
As expected for \ac{ISC}, the effect of resummation is small because there cannot be an enhancement by resonances.
Hence, these corrections are standard \ac{NNNLO} terms of the order of $10^{-6}$ relative to the \ac{LO}.
The \ac{NLO} \ac{ISC} corrections described by~\eqref{eq:NLOISCimpact} are around~$10\%$, as already observed in~\cite{Aliberti:2024fpq}.

The $\delta$ defined in~\eqref{eq:thedelta} illustrates the importance and symmetry of \ac{NLO} mixed corrections compared to \ac{NNLO} \ac{ISC}.
The \ac{NLO} mixed corrections have a strong asymmetry that depends on the pion description used, and exceed the size of the symmetric \ac{NNLO} \ac{ISC} in the bulk of the distributions. Only at kinematic boundaries, as e.g.~illustrated  with $|\cos\theta^+|\simeq 0.55$  in Figure~\ref{fig:costh}, the \ac{NNLO} terms are of similar size. 

\begin{figure}
    \centering
    \includegraphics{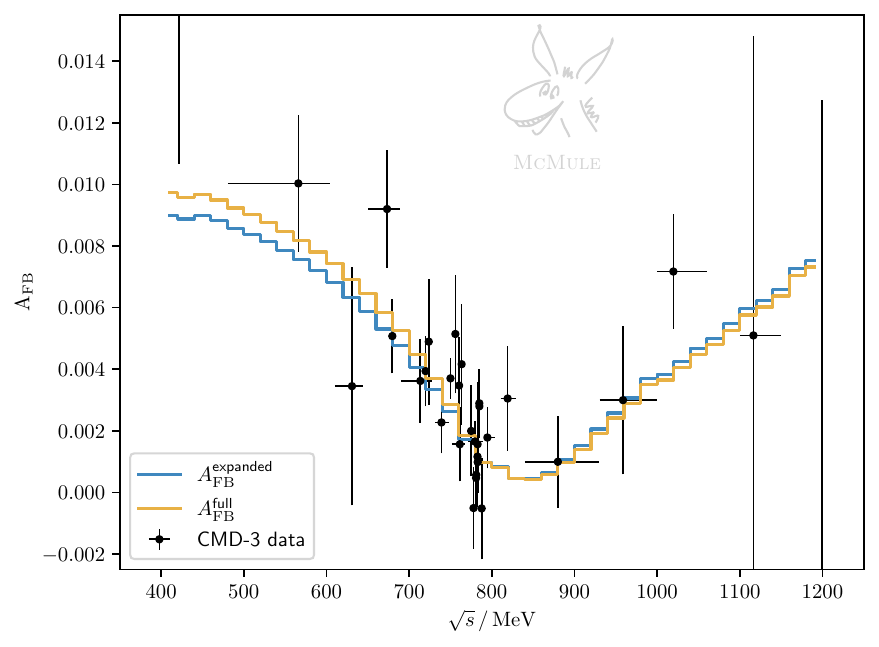}
    \caption{Different versions of the forward-backward charge asymmetry defined in~\eqref{eq:defAFBversions} together with CMD-3 data.
    }
    \label{fig:asym}
\end{figure}

Figure~\ref{fig:asym} shows the asymmetry as a function of $\sqrt{s}$ in the energy range of the CMD-3 experiment, i.e.
\begin{align}
\label{eq:CMDasym}
    A_{\rm FB}(\sqrt{s}) = \frac{\sigma(\sqrt{s})\vert_{\thavg>\frac{\pi}{2}} - \sigma(\sqrt{s})\vert_{\thavg<\frac{\pi}{2}}}{\sigma(\sqrt{s})\vert_{\thavg>\frac{\pi}{2}} + \sigma(\sqrt{s})\vert_{\thavg<\frac{\pi}{2}}}\,.
\end{align}
As illustration, the CMD-3 measurements~\cite{CMD-3:2023alj,CMD-3:2023rfe} are also shown in Figure~\ref{fig:asym}. This quantity is completely driven by the mixed corrections and the \ac{NNLO} \ac{ISC} contribute only to the denominator.
The absence of the $\rho/\phi$ interference at $\sqrt{s}\simeq m_\phi$ in our results is due to its absence in the \ac{VFF} that only considered CMD-2 data~\cite{Colangelo:2018mtw}, where this interference was not visible. 
When expanding~\eqref{eq:CMDasym} in terms of $\alpha$, we can define two versions
\begin{subequations}
\label{eq:defAFBversions}
    \begin{align}
    A^{\rm expanded}_{\rm FB}(\sqrt{s}) &= \frac{\sigma_1(\sqrt{s})\vert_{\thavg>\frac{\pi}{2}} - \sigma_1(\sqrt{s})\vert_{\thavg<\frac{\pi}{2}}}{\sigma_0(\sqrt{s})\vert_{\thavg>\frac{\pi}{2}} + \sigma_0(\sqrt{s})\vert_{\thavg<\frac{\pi}{2}}} + \mathcal{O}(\alpha^2)\,, \\
    A^{\rm full}_{\rm FB}(\sqrt{s}) &= \frac{\sigma_1(\sqrt{s})\vert_{\thavg>\frac{\pi}{2}} - \sigma_1(\sqrt{s})\vert_{\thavg<\frac{\pi}{2}}}{\sigma_1(\sqrt{s})\vert_{\thavg>\frac{\pi}{2}} + \sigma_1(\sqrt{s})\vert_{\thavg<\frac{\pi}{2}}}\,.
\end{align}
\end{subequations}
While the first version, $A^{\rm expanded}_{\rm FB}(\sqrt{s})$, is a strict expansion in $\alpha$, the second version, $ A^{\rm full}_{\rm FB}(\sqrt{s})$, uses the \ac{NLO} prediction everywhere.
The difference between these versions is the overall normalisation in the denominator. 
As visible in Figure~\ref{fig:asym}, this does have an impact, in particular for small values of $\sqrt{s}$.
Except for deviations due to different choices of the \ac{VFF}, our result for $A^{\rm full}_{\rm FB}$ agrees with~\cite{Budassi:2024whw}.

\section {Conclusion and outlook} \label{sec:conclusion}

We have presented a method, disperon QED, to incorporate data input into a Monte Carlo for loop processes.
It is based on treating the effects of a dispersion relation through a new particle, the disperon, which allows us to rely on automated tools such as \OpenLoops{}.
Numerical challenges arising from the dispersive integration are dealt with using an \ac{EFT}-inspired approach, \ac{DET}.
This, combined with a universal threshold subtraction, presents a method with a wide range of applicability. 

The method has been applied to the process \eepp{} and implemented in \mcmule. While similar results  for this process are already available in the literature, the main thrust of this article is to prepare for applications to more complicated processes. 

We plan to apply disperon QED to \eeppg{}, initially considering mixed corrections with photon emission from the electron line. 
Adding QED emission to the initial state does not affect the applicability of \ac{FsQED} for the hadronic part.
In fact, it is even conceivable to go one order higher on this part of the amplitudes.
In a second step we will include mixed-correction contributions with photon emission from the final state for \eeppg.
Even in this case, \ac{FsQED} is a reasonable starting point.  Similarly, form-factor-improved interactions of a photon with protons can be used as a model to describe aspects of lepton-proton scattering. While this requires a slight extension with additional disperon couplings, the core ideas are the same.

Finally, we have considered the resummation of \ac{VP} contributions within loops. These resummation effects are beyond \ac{NNLO} and, hence, typically very small. However, in the vicinity of resonance structures, there can be a substantial enhancement. It will be interesting to investigate these effects for $2\to{2}$ scattering in kinematic situations where e.g.~the $J/\psi$ resonance enters.

With the methods developed here, the technical challenges related to a multitude of phenomenologically relevant computations with generic numerical input in loops can be tackled. Rather than addressing them on a case-by-case base, our approach exploits the availability of general tools and concepts, and provides a versatile tool that renders many computations feasible.

\subsection*{Acknowledgements}
It is a pleasure to thank Martin Hoferichter for providing the pion VFF used in this work, for discussions about it, and for clarifying the applicability of various approaches for hadronic matrix elements. We are grateful to Fedor Ignatov and Andrea Gurgone for providing the $A_{\rm FB}$ data measured at CMD-3 and calculated in~\cite{Budassi:2024whw}, respectively, to compare with our results. We are also grateful to Sumit Banik for discussions about Landau singularities, as well as to Martina Cottini and Simon Holz for discussions about threshold subtraction in $\tau$ decays and for comparing results of \ac{FsQED} in dimensional regularisation. We wish to thank Franziska Hagelstein and Matteo Ronchi for discussions about the applicability of disperon QED in lepton-proton scattering, as well as for their input to Appendix~\ref{sec:ComptonTensor}.

SK acknowledges support by the Swiss National Science Foundation (SNSF) under grant 207386.
MR has been supported by the Italian Ministry of Universities and Research (MUR) through grants PRIN 2022BCXSW9 and FIS (CUP: D53C24005480001, FLAME).
The research of MZ was supported by the SNSF under the contract TMSGI2-211209.

\appendix

\section{Derivation of the dispersion relation}\label{sec:disp-proof}

\begin{figure}
    \centering
    \begin{tikzpicture}
        \def\delta{0.1}
        \def\th{0.5}
        \def\R{2}
        \def\theta{atan(\delta/\R)}
        \draw[->](-2.3,0) -- (2.3,0);
        \draw[->](0,-2.3) -- (0,2.3);
        
        \begin{scope}[decoration={markings,mark=at position 0.5 with {\arrow{>}}}]
            \draw[postaction={decorate}] (\th,\delta) -- (\R,\delta);
            \centerarc[](\th,0)(270:90:\delta);
            \centerarc[postaction={decorate}](0,0)({\theta}:{360-\theta}:\R);
            \draw[postaction={decorate}] (\R,-\delta) -- (\th,-\delta);
        \end{scope}
    \end{tikzpicture}
    \caption{The contour $\mathcal{C}$ for the dispersion relation.}
    \label{fig:contour}
\end{figure}
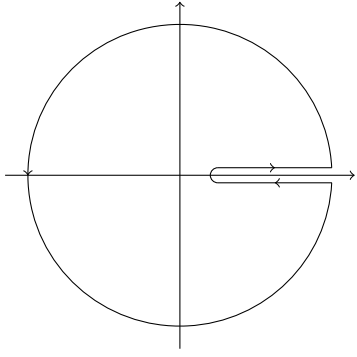

For completeness we repeat the derivation of the dispersion relation for a function $F(s)$ which will help us understand what requirements this function needs to fulfil. Our discussion follows closely~\cite{Colangelo:2025sah,Jegerlehner:2017gek,Kniehl:1996rh}. 
To start with, 
$F(s)$ shall be analytic in the $s$-plane except for a cut along the positive real axis starting at some finite $s=\sth$.
This means that Cauchy's integral formula is applicable
\begin{align}
    F(s+i\delta) 
        = \frac{1}{2\pi i} \oint_{\mathcal{C}} \frac{F(\dpone)}{\dpone-s} \D \dpone
        = \frac{1}{2\pi i} \Bigg(
            \int_{\sth+i\delta}^{R+i\delta} + \int_\text{large arc} - \int_{\sth-i\delta}^{R-i\delta} + \int_\text{small arc}
          \Bigg) \frac{F(\dpone)}{\dpone-s} \D \dpone\,,
\end{align}
where the integration contour $\mathcal{C}$ is defined as shown in Figure~\ref{fig:contour} and $s$ is inside the contour.
Assuming the function is well-behaved around threshold (for example by being linear or constant), the integral over the small arc vanishes as $\delta\to0$.
Similarly, if $F$ falls off fast enough ($\dpone^{-1-\epsilon}$ for $\epsilon > 0$) as $|\dpone|\to\infty$, the large arc vanishes as well.
Therefore,
\begin{align}
    F(s) = \lim_{\delta\to0^+} F(s+i\delta) 
        = \lim_{\delta\to0^+}\frac{1}{2\pi i} \Bigg(
            \int_{\sth+i\delta}^{\infty} - \int_{\sth-i\delta}^{\infty}
          \Bigg) \frac{F(\dpone)}{\dpone-s} \D \dpone\,.
\end{align}
We can combine the two integrals
\begin{align}
    F(s)
       = \lim_{\delta\to0^+} \frac1{2\pi i} \int_\sth^\infty \frac{F(\dpone+i\delta)-F(\dpone-i\delta)}{\dpone-s-i\delta} \D \dpone
       = \lim_{\delta\to0^+} \frac1{\pi} \int_\sth^\infty \frac{\Im \, F(\dpone)}{\dpone-s-i\delta} \D \dpone\, .
\end{align}
In the last step we have used the Schwarz reflection principle
\begin{align}
F(s^*) = F^*(s)\quad \forall\, s\, \text{except the cut}\,,
\end{align}
which requires $F(s)$ to be real for real-valued $s$ below the threshold.
Hence, $F(\dpone+i\delta)-F(\dpone-i\delta) = 2i\,\Im\,F(\dpone)$ for $\delta\to0$.
However, the functions we consider often do not fall off quickly enough, i.e.~$F(s)$ does not scale like $s^{-1-\epsilon}$.
Instead, we use a (once) subtracted dispersion relation and decompose $F(s) \to (F(s)-F(0))/s$
\begin{align}
    \frac{F(s)-F(0)}{s} = \lim_{\delta\to0^+} \frac1{\pi} \int_\sth^\infty \frac{\Im \,F(\dpone) - \Im\,F(0)}{\dpone-s-i\delta} \frac{\D \dpone}{\dpone}\,.
\end{align}
Often, we have $F(0)=0$ (e.g.~in the case of \ac{VP} corrections) or $F(0)=1$ (in the case of form factors).
Under this assumption we can further simplify
\begin{align}
    \frac{F(s) - F(0)}{s} = - \lim_{\delta\to0^+} \frac1{\pi} \int_\sth^\infty \frac{\Im \, F(\dpone)}{s-\dpone+i\delta} \frac{\D \dpone}\dpone\,.
\end{align}

In summary, the (once subtracted) dispersion relation requires that $F(s)$ is analytic except for $s>\sth$ along the real axis, is real for real-valued $s$ below the threshold, falls off faster than $F\sim s^{-\epsilon}$ for any $\epsilon>0$, and is well-behaved around the threshold.

\section{Compton tensor} \label{sec:ComptonTensor}

In this appendix we will briefly summarise the background of \ac{FsQED} and \ac{FxsQED}.
Our discussion will follow~\cite{Colangelo:2015ama,Colangelo:2017fiz,Hoferichter:2019nlq} and its aim is to clarify the notation and the relationship between \ac{FsQED} and \ac{FxsQED}.

In \eqref{eq:feynmanrule:fxsqed} we have stated the Feynman rule for the process $\gamma^*(k)\to \pi^+(p)\pi^-(q)$ as a \ac{sQED} Feynman rule multiplied with a form factor, i.e.~\ac{FxsQED}.
It is not clear that this would immediately imply that one-loop calculations as in~\eqref{eq:A1mixed} can be carried out using the same Feynman rule.
Instead, we need to begin by studying the full doubly-virtual Compton tensor for $\gamma^*(k_1)\gamma^*(k_2) \to \pi^+(p)\pi^-(q)$
\begin{align}
    W^{\mu\nu}(p, q; k_1,k_2) = \int\D^4x\ e^{-i k_1\cdot x} \langle \pi^+(p) \pi^-(q) \vert j_{em}^\mu(0)j_{em}^\nu(x)\vert 0 \rangle
    =
    \begin{gathered}
        \begin{tikzpicture}
            \draw[dashed] (0, 0.5) -- (  45:1) node [right] {$\pi^+(p)$};
            \draw[dashed] (0,-0.5) -- ( -45:1) node [right] {$\pi^-(q)$};
            \draw[photon] (0, 0.5) -- ( 135:1) node [left ] {$\gamma^*(k_1)$};
            \draw[photon] (0,-0.5) -- (-135:1) node [left ] {$\gamma^*(k_2)$};
            \fill[gray] (0,0.) ellipse (0.3 and 0.7);
        \end{tikzpicture}
    \end{gathered}\,.
\end{align}
The full tensor is a complicated, non-perturbative object.
However, it can be written in terms of five form factors that can in principle be measured or estimated~\cite{Colangelo:2015ama,Colangelo:2017fiz,Hoferichter:2019nlq}.
Alternatively, we can derive a dispersive decomposition of $W^{\mu\nu}$.
To do this, one fixes e.g.~$s=(p+q)^2$ and considers the discontinuities $D_i$ along the $t=(k_1-p)^2$ and $u=(k_2-p)^2$ cuts
\begin{align}
    W^{\mu\nu} = \frac1\pi\int_0^\infty\D t'\frac{D_t^{\mu\nu}(t',s)}{t'-t} + \frac1\pi\int_0^\infty\D u'\frac{D_u^{\mu\nu}(u',s)}{u'-u}\,.
\end{align}
Using unitarity relations, the discontinuities can be written as a sum over all possible intermediary states $|n\rangle$.
Schematically,
\begin{align}
    D_t(t,s)\sim\sum_{n}\int\D\Phi_n \langle \pi^-(q)\gamma^*(-k_2) | n\rangle \langle n | \pi^-(-p)\gamma^*(k_1)\rangle\,.
\end{align}
The first state in this sum is the pion itself, $|n\rangle = |\pi\rangle$, which results in (the dispersive cut is indicated with the black line crossing the pion propagator)
\begin{align}
    D_t^{\mu\nu}(t,s) &=
        \begin{gathered}
            \begin{tikzpicture}
                \draw[dashed] (0, 0.5) -- (  45:1) node [right] {$\pi^+(p)$};
                \draw[dashed] (0,-0.5) -- ( -45:1) node [right] {$\pi^-(q)$};
                \draw[photon] (0, 0.5) -- ( 135:1) node [left ] {$\gamma^*(k_1)$};
                \draw[photon] (0,-0.5) -- (-135:1) node [left ] {$\gamma^*(k_2)$};
                \draw[dashed] (0, 0.5) -- (0,-0.5);
                \fill[black] (0, 0.5) circle (0.08);
                \fill[black] (0,-0.5) circle (0.08);
                \draw (-0.2,0) -- (0.2,0);
            \end{tikzpicture}
        \end{gathered}\\
    &=
    e^4\pi \delta(t-m_\pi^2) F_\pi^V(k_1^2) F_\pi^V(k_2^2)\Bigg(
        k_1^\nu k_2^\mu
        -\frac12 (s-k_1^2-k_2^2) g^{\mu\nu}
        -Q^\mu Q^\nu\notag\\&\qquad\qquad
        +2k_1\cdot Q \frac{k_2^\mu Q^\nu-k_1^\nu Q^\mu+k_1\cdot Q g^{\mu\nu}}{s-k_1^2-k_2^2}
    \Bigg)\,, \nonumber
\end{align}
with $Q=p-q$ and $D_u$ with $t\leftrightarrow u$.
Therefore, we can write $W^{\mu\nu}$ as
\begin{align}
    W^{\mu\nu} &=
     e^4F_\pi^V(k_1^2) F_\pi^V(k_2^2)\ \Bigg(
        \frac{(2 k_1^\mu-p^\mu) (2 k_2^\nu-q^\nu)}{t-m_\pi^2}
       +\frac{(2 k_1^\nu-q^\nu) (2 k_2^\mu-p^\mu)}{u-m_\pi^2}+2  g^{\mu\nu}\Bigg) \nonumber
    \\&=
        \begin{gathered}
            \begin{tikzpicture}
                \draw[dashed] (0, 0.5) -- (  45:1);% node [right] {$\pi^+(p)$};
                \draw[dashed] (0,-0.5) -- ( -45:1);% node [right] {$\pi^-(q)$};
                \draw[photon] (0, 0.5) -- ( 135:1);% node [left ] {$\gamma^*(k_1)$};
                \draw[photon] (0,-0.5) -- (-135:1);% node [left ] {$\gamma^*(k_2)$};
                \draw[dashed] (0, 0.5) -- (0,-0.5);
                \fill[black] (0, 0.5) circle (0.08);
                \fill[black] (0,-0.5) circle (0.08);
                \draw (-0.2,0) -- (0.2,0);
            \end{tikzpicture}
        \end{gathered}+
        \begin{gathered}
            \begin{tikzpicture}
                \draw[dashed] (0, 0.5) -- ( -45:1);% node [right] {$\pi^+(p)$};
                \draw[dashed] (0,-0.5) -- (  45:1);% node [right] {$\pi^-(q)$};
                \draw[photon] (0, 0.5) -- ( 135:1);% node [left ] {$\gamma^*(k_1)$};
                \draw[photon] (0,-0.5) -- (-135:1);% node [left ] {$\gamma^*(k_2)$};
                \draw[dashed] (0, 0.5) -- (0,-0.5);
                \fill[black] (0, 0.5) circle (0.08);
                \fill[black] (0,-0.5) circle (0.08);
                \draw (-0.2,0) -- (0.2,0);
            \end{tikzpicture}
        \end{gathered}
    = F_\pi^V(k_1^2) F_\pi^V(k_2^2)\Bigg(
    \begin{gathered}
        \begin{tikzpicture}
            \draw[dashed] (0, 0.5) -- (  45:1);% node [right] {$\pi^+(p)$};
            \draw[dashed] (0,-0.5) -- ( -45:1);% node [right] {$\pi^-(q)$};
            \draw[photon] (0, 0.5) -- ( 135:1);% node [left ] {$\gamma^*(k_1)$};
            \draw[photon] (0,-0.5) -- (-135:1);% node [left ] {$\gamma^*(k_2)$};
            \draw[dashed] (0, 0.5) -- (0,-0.5);
        \end{tikzpicture}
    \end{gathered}+
    \begin{gathered}
        \begin{tikzpicture}
            \draw[dashed] (0, 0.5) -- ( -45:1);% node [right] {$\pi^+(p)$};
            \draw[dashed] (0,-0.5) -- (  45:1);% node [right] {$\pi^-(q)$};
            \draw[photon] (0, 0.5) -- ( 135:1);% node [left ] {$\gamma^*(k_1)$};
            \draw[photon] (0,-0.5) -- (-135:1);% node [left ] {$\gamma^*(k_2)$};
            \draw[dashed] (0, 0.5) -- (0,-0.5);
        \end{tikzpicture}
    \end{gathered}+
    \begin{gathered}
        \begin{tikzpicture}
            \draw[dashed] (0,0) -- ( -45:1);% node [right] {$\pi^+(p)$};
            \draw[dashed] (0,0) -- (  45:1);% node [right] {$\pi^-(q)$};
            \draw[photon] (0,0) -- ( 135:1);% node [left ] {$\gamma^*(k_1)$};
            \draw[photon] (0,0) -- (-135:1);% node [left ] {$\gamma^*(k_2)$};
        \end{tikzpicture}
    \end{gathered}\Bigg)\,.
    \label{eq:compton:bornpole}
\end{align}
In other words, it is possible to obtain the pion-pole contribution ($|n\rangle=|\pi\rangle$) for $\gamma^*\gamma^*\to\pi\pi$ by calculating in \ac{sQED} and multiplying with the \acp{VFF}, i.e. \ac{FxsQED}.

Note that the pion-pole approximation used here assumes that the propagator of the intermediary particle still remains that of a pion.
This approximation has been shown to be good enough for $\gamma^*\gamma^*\to\pi\pi$ at low energies.
For $\gamma^*\gamma^*\gamma\to\pi\pi$, it is less clear whether \ac{FxsQED} is a good approximation. 
A first step has been taken by constructing the dispersive decomposition of this new Compton tensor in the limit where the on-shell photon is soft~\cite{Kaziukenas:2025gggpp}.
This has indicated that \ac{FxsQED} for the Compton tensor is a well-defined reasonable starting point which was also indicated via the so-called Pisa Consensus of the RadioMonteCarLow 2 Group~\cite{Rocco:2025TI}.

Beyond pions, a similar construction is possible for $\gamma^*\gamma^*\to pp$ and $\gamma^*p\to \gamma^*p$ as entering $\ell p\to \ell p$ scattering.
However, the equivalent relation of~\eqref{eq:compton:bornpole} for a spinor instead of a scalar no longer holds:
the proton-pole contribution no longer agrees with the Born amplitudes that can be calculated using the equivalent of \ac{FxsQED} where we use a Feynman rule that involves form factors (now two rather than just a single \ac{VFF})
\begin{align}
    \begin{gathered}
        \begin{tikzpicture}
            \draw (-0.5,0) node [left] {$p(p)$} -- (0.5,0) node [right] {$p(q)$};
            \draw[photon] (0,0) -- (0,1) node [right] {$\gamma^*(k)$};
            \fill[gray] (0,0) circle (0.2);
        \end{tikzpicture}
    \end{gathered}
    = F_1(k^2)\ \gamma^\mu + F_2(k^2)\ \frac{i}{2M} \ \sigma^{\mu\nu} k_\nu\,.
\end{align}
This is because the Born term includes contributions that are not due to the proton pole which need to be subtracted back out.
Schematically,
\begin{align}
    \begin{gathered}
        \begin{tikzpicture}[rotate=270]
            \draw         (0, 0.5) -- (  45:1) node [right] {$q$};% node [right] {$\pi^+(p)$};
            \draw         (0,-0.5) -- ( -45:1) node [left ] {$p$};% node [right] {$\pi^-(q)$};
            \draw[photon] (0, 0.5) -- ( 135:1) node [right] {$k_2$};% node [left ] {$\gamma^*(k_1)$};
            \draw[photon] (0,-0.5) -- (-135:1) node [left ] {$k_1$};% node [left ] {$\gamma^*(k_2)$};
            \draw         (0, 0.5) -- (0,-0.5);
            \fill[black] (0, 0.5) circle (0.08);
            \fill[black] (0,-0.5) circle (0.08);
            \draw (-0.2,0) -- (0.2,0);
        \end{tikzpicture}
    \end{gathered}+
    \begin{gathered}
        \begin{tikzpicture}[rotate=270]
            \draw         (0, 0.5) -- ( -45:1);% node [right] {$\pi^+(p)$};
            \draw         (0,-0.5) -- (  45:1);% node [right] {$\pi^-(q)$};
            \draw[photon] (0, 0.5) -- ( 135:1);% node [left ] {$\gamma^*(k_1)$};
            \draw[photon] (0,-0.5) -- (-135:1);% node [left ] {$\gamma^*(k_2)$};
            \draw         (0, 0.5) -- (0,-0.5);
            \fill[black] (0, 0.5) circle (0.08);
            \fill[black] (0,-0.5) circle (0.08);
            \draw (-0.2,0) -- (0.2,0);
        \end{tikzpicture}
    \end{gathered}
    =
    \begin{gathered}
        \begin{tikzpicture}[rotate=270]
            \draw         (0, 0.5) -- (  45:1);% node [right] {$\pi^+(p)$};
            \draw         (0,-0.5) -- ( -45:1);% node [right] {$\pi^-(q)$};
            \draw[photon] (0, 0.5) -- ( 135:1);% node [left ] {$\gamma^*(k_1)$};
            \draw[photon] (0,-0.5) -- (-135:1);% node [left ] {$\gamma^*(k_2)$};
            \draw         (0, 0.5) -- (0,-0.5);
            \fill[gray] (0, 0.5) circle (0.2);
            \fill[gray] (0,-0.5) circle (0.2);
        \end{tikzpicture}
    \end{gathered}+
    \begin{gathered}
        \begin{tikzpicture}[rotate=270]
            \draw         (0, 0.5) -- ( -45:1);% node [right] {$\pi^+(p)$};
            \draw         (0,-0.5) -- (  45:1);% node [right] {$\pi^-(q)$};
            \draw[photon] (0, 0.5) -- ( 135:1);% node [left ] {$\gamma^*(k_1)$};
            \draw[photon] (0,-0.5) -- (-135:1);% node [left ] {$\gamma^*(k_2)$};
            \draw         (0, 0.5) -- (0,-0.5);
            \fill[gray] (0, 0.5) circle (0.2);
            \fill[gray] (0,-0.5) circle (0.2);
        \end{tikzpicture}
    \end{gathered}
    - \sum_{i=1}^2 F_i(k_1^2)F_i(k_2^2) T_i^{\mu\nu}(p,q;k_1,k_2)\,,
\end{align}
with two tensor functions $T_i^{\mu\nu}$ that can be reconstructed from~\cite{Eichmann:2018ytt,Hagelstein:2017cbl}
\begin{align}
    T_1^{\mu\nu} & = \frac{1}{M} g^{\mu\nu}\,,\\
    T_2^{\mu\nu} & = \frac1{4iM^3}\Big(-i\ k_1^\mu k_2^\nu + k_1\cdot(p+q)\ \sigma^{\mu\nu} + \gamma^5 \varepsilon^{\mu\nu\alpha\beta}k_{2\alpha}k_{1\beta}\Big)\,.
\end{align}
For a discussion on the difference between Born and pole contributions, see for example~\cite{Scherer:1996ux,Drechsel:1997xv,Birse:2012eb,Gasser:2015dwa}.
In the case of forward scattering ($(p-k_1)^2\sim0$), we also refer to the review in~\cite{Hagelstein:2015egb}.

Both for pions and protons, contributions beyond the pion or proton pole are delicate.
In the pion case, the intermediary vector-meson states need to be unitarised to get the $f_2(1270)$ resonance correct which destroys the simple \ac{FxsQED} picture for the Compton tensor~\cite{Hoferichter:2019nlq}.
For the proton case the pole approximation for the $\Delta(1232)$ intermediate state has been shown to be insufficient~\cite{Hagelstein:2018bdi}.

\section{Renormalisation}
\label{sec:renormalisation}
Including a disperon of mass $\sqrt{\dpone}$ does add additional terms to the QED renormalisation constants.
In total, the vertex and mass renormalisation read
\begin{subequations}
\label{eq:changedrenormconstants}
\begin{alignat}{2}
    \delta Z^{(1)}_2(\sqrt{\dpone}) & = \frac{\alpha}{2\pi} \left(\frac{\mu^2}{m^2}\right)^\epsilon && \left[
        - 2 - \frac{3}{2 \epsilon} \right. \nonumber\\
        & &&-\frac{3\left(4m^4 + 2m^2 \dpone - \dponesq\right)}{2 m^2 (4 m^2 - \dpone)} \, \text{DiscB}(m^2,m,\sqrt{\dpone})  \\
       & &&+  \left.\left(1-\frac{3 \dponesq}{4 m^4}\right) \log\left(\frac{m^2}{\dpone}\right)
        - 2 - \frac{1}{2 \epsilon} - \frac{3 \dpone}{2 m^2} 
    \right]\,, \nonumber\\
    \delta Z^{(1)}_m (\sqrt{\dpone}) & = \frac{\alpha}{2\pi} \left(\frac{\mu^2}{m^2}\right)^\epsilon && \left[ - 2 - \frac{3}{2 \epsilon} \right. \nonumber\\
        & && -\frac{1}{4}\left(4+\frac{2\dpone}{m^2}\right)\text{DiscB}(m^2,m,\sqrt{\dpone})  \\
    & && -\left.\frac{\dponesq}{4 m^4} \log\left(\frac{m^2}{\dpone}\right) - \frac{1}{4} \left(8 + \frac{6}{\epsilon}+ \frac{2 \dpone}{m^2} \right) \right]\,, \nonumber
\end{alignat}
\end{subequations}
where we have used from {\tt Package-X}
\begin{align}
\text{DiscB}(m^2,m,\sqrt{\dpone}) = \frac{\sqrt{\dpone\left(\dpone-4 m^2\right)}}{m^2} \log\left(\frac{\dpone+\sqrt{\dpone\left(\dpone-4 m^2\right)}}{2 m \sqrt{\dpone}}\right)\,.
\end{align}
The first lines of~\eqref{eq:changedrenormconstants} contain the pure QED contribution.
Note that~\eqref{eq:changedrenormconstants} are given in the 't~Hooft--Veltman scheme.
As discussed in Section~\ref{sec:openloops}, they need to be combined with the rational \acp{CT} before being passed to \OpenLoops{}.

\section{Feynman rules of DET}
\label{sec:feynmanrules}
Here, we collect the Feynman rules for the operators in the DET Lagrangian~\eqref{eq:Lmatch}.
We include rules for the process $e^-(p_1)e^+(p_2)(\ell^+(p_3) \ell^-(p_4))\to\pi^-(q_1)\pi^+(q_2)(\gamma(q_3))$
\begin{alignat}{2}
     \fourpointsixd{1}\quad&=\quad -i (\gamma_\mu)(q_1-q_2)^\mu\,,
    &&\fivepointsixd{1}\quad=\quad 2ie(\gamma_\mu)\,,\\
     \fourpointeightd{1}\quad&=\quad -i (p_1+p_2)^2 \,(\gamma_\mu)(q_1-q_2)^\mu\,,  \quad 
     &&\fivepointeightd{1}\quad=\quad 2ie\,(p_1+p_2)^2\,(\gamma_\mu)\,, \\
     \sixpoint{1}\quad&=\quad i\,(\gamma_\mu)[\gamma^\mu]\,,
\end{alignat}
as well as for $e^-(p_1)e^+(p_2)\to\ell^-(q_1)\ell^+(q_2)$
\begin{align}
        \hvpsix{1}\quad&=\quad i\,(\gamma_\mu)[\gamma^\mu]\,, \\
     \hvpeight{1}\quad&=\quad -i\,(p_1+p_2)^2 (\gamma_\mu)[\gamma^\mu]\,.
\end{align}
This process specification means that the momenta $p_i$ are considered as incoming, the momenta $q_i$ as outgoing.
Additionally, the Wilson coefficients $C_i$ have to be included in the Feynman rule. 
This was for example done in~\eqref{eq:examplesoft}.

\bibliographystyle{JHEP}
\bibliography{pions}

\end{document}

%% file: orcidlink.tex
\usetikzlibrary{svg.path}

\definecolor{orcidlogocol}{HTML}{A6CE39}
\newcommand\orcidlink[1]{\href{https://orcid.org/#1}{%
\begin{tikzpicture}[yscale=-0.026,xscale=0.026,transform shape]
    \fill[orcidlogocol] svg{M256,128c0,70.7-57.3,128-128,128C57.3,256,0,198.7,0,128C0,57.3,57.3,0,128,0C198.7,0,256,57.3,256,128z};
    \fill[white] svg{M86.3,186.2H70.9V79.1h15.4v48.4V186.2z}
                 svg{M108.9,79.1h41.6c39.6,0,57,28.3,57,53.6c0,27.5-21.5,53.6-56.8,53.6h-41.8V79.1z M124.3,172.4h24.5c34.9,0,42.9-26.5,42.9-39.7c0-21.5-13.7-39.7-43.7-39.7h-23.7V172.4z}
                 svg{M88.7,56.8c0,5.5-4.5,10.1-10.1,10.1c-5.6,0-10.1-4.6-10.1-10.1c0-5.6,4.5-10.1,10.1-10.1C84.2,46.7,88.7,51.3,88.7,56.8z};
\end{tikzpicture}\!}%
}

%% file: figures/tikz-diags.tex
\def\centerarc[#1](#2,#3)(#4:#5:#6)% Syntax: [draw options] (center) (initial angle:final angle:radius)
    { \draw[#1] ({#2+#6*cos(#4)},{#3+#6*sin(#4)}) arc (#4:#5:#6); }

\newcommand\boxdiag[2]{
    \begin{gathered}
        \begin{tikzpicture}
            \draw (135:0.7) -- (135:0.5) -- (-135:0.5) -- (-135:0.7);
            \draw (45:0.7) -- (45:0.5) -- (-45:0.5) -- (-45:0.7);
            \draw[#1] (135:0.5) -- (45:0.5);
            \draw[#2] (-135:0.5) -- (-45:0.5);
        \end{tikzpicture}
    \end{gathered}
}
\newcommand\boxdiagphoton[2]{
        \begin{tikzpicture}[baseline=-0.5ex]
            \draw (135:0.7) -- (135:0.5) -- (-135:0.5) -- (-135:0.7);
            \draw (45:0.7) -- (45:0.5) -- (-45:0.5) -- (-45:0.7);
            \draw[#1] (135:0.5) -- (45:0.5);
            \draw[#2] (-135:0.5) -- (-45:0.5);
            \draw[photon] (-135:0.6) -- ++(-40:0.5);
        \end{tikzpicture}
}
\newcommand\boxdiagphotontwo[2]{
        \begin{tikzpicture}[baseline=-0.5ex]
            \draw (135:0.7) -- (135:0.5) -- (-135:0.5) -- (-135:0.7);
            \draw (45:0.7) -- (45:0.5) -- (-45:0.5) -- (-45:0.7);
            \draw[#1] (135:0.5) -- (45:0.5);
            \draw[#2] (-135:0.5) -- (-45:0.5);
            \draw[photon] (135:0.6) -- ++(40:0.5);
        \end{tikzpicture}
}
\newcommand\pentagon[2]{
        \begin{tikzpicture}[baseline=-0.5ex]
            \draw (135:0.7) -- (135:0.5) -- (-135:0.5) -- (-135:0.7);
            \draw (45:0.7) -- (45:0.5) -- (-45:0.5) -- (-45:0.7);
            \draw[#1] (135:0.5) -- (45:0.5);
            \draw[#2] (-135:0.5) -- (-45:0.5);
            \draw[photon] (-0.35, 0) -- ++(0:-0.4);
        \end{tikzpicture}
}

\def\disperonphotonbox{\boxdiag{zigzag, line width=1.2pt}{photon}}

\newcommand{\seagulld}[1]{
    \begin{tikzpicture}[scale=#1,baseline=-0.5ex]
    \draw (0,-1) -- (0.8,-1) -- (0.8,1) -- (0,1); 
    \draw[zigzag, line width=1.2pt] (0.8,-1) -- (2.4,0);
    \draw[zigzag, line width=1.2pt] (0.8,1) -- (2.4,0);
    \draw[dashed] (3.6,-1) -- (2.4,0) -- (3.6,1);
    \end{tikzpicture}
}

\newcommand{\seagulleftdd}[1]{
    \begin{tikzpicture}[scale=#1,baseline=-0.5ex]
    \draw[dashed] (45:0) -- (45:1.2);
    \draw (-135:0) -- (-135:1.2);
    \draw (135:0) -- (135:1.2);
    \draw[dashed] (-45:0.22) -- (-45:1.2);
    \draw
        (0,0) .. controls ++(-1.2,0.8) and ++(-1.2,-0.8) .. (0,0);
    \draw[fill=white] (0:0) circle (0.22);
    \node at (0:0) {3};
    \end{tikzpicture}
}
\newcommand{\seagulls}[1]{
    \begin{tikzpicture}[scale=#1,baseline=-0.5ex]
    \draw (0,-1) -- (0.8,-1) -- (0.8,1) -- (0,1); 
    \draw[zigzag, line width=1.2pt] (0.8,-1) -- (2.4,0);
    \draw[photon] (0.8,1) -- (2.4,0);
    \draw[dashed] (3.6,-1) -- (2.4,0) -- (3.6,1);
    \end{tikzpicture}
}

\newcommand{\seagulleftssix}[1]{
    \begin{tikzpicture}[scale=#1,baseline=-0.5ex]
    \draw (135:0.2) -- (135:1.2);
    \draw  (-135:0.2) -- (-135:1.2);
    \draw[dashed] (45:0.2) -- (45:1.2);
    \draw[dashed] (-45:0.2) -- (-45:1.2);
    \draw[photon] (-135:0.9) arc[start angle=-135, end angle=20, radius=0.4]; 
    \draw (0:0) circle (0.22);
    \node at (0:0) {1};
    \end{tikzpicture}
}
\newcommand{\seagulleftseight}[1]{
    \begin{tikzpicture}[scale=#1,baseline=-0.5ex]
    \draw (135:0.2) -- (135:1.2);
    \draw  (-135:0.2) -- (-135:1.2);
    \draw[dashed] (45:0.2) -- (45:1.2);
    \draw[dashed] (-45:0.2) -- (-45:1.2);
    \draw[photon] (-135:0.9) arc[start angle=-135, end angle=20, radius=0.4]; 
    \draw (0:0) circle (0.22);
    \node at (0:0) {2};
    \end{tikzpicture}
}

\tikzset{
  midarrow/.style={
    postaction={decorate},
    decoration={markings, mark=at position 0.5 with {\arrow{<}}}
  }
}

\newcommand{\fourpointsixd}[1]{
    \begin{tikzpicture}[scale=#1,baseline=-0.5ex]
    \draw[midarrow] (135:0.2) -- (135:1);
    \draw[midarrow] (-135:1) -- (-135:0.2);
    \draw[dashed] (45:0.2) -- (45:1);
    \draw[dashed] (-45:0.2) -- (-45:1);
    \fill[white] (0:0) circle (0.22);
    \draw (0:0) circle (0.22);
    \node at (0:0) {1};
    \end{tikzpicture}
}
\newcommand{\fivepointsixd}[1]{
    \begin{tikzpicture}[scale=#1,baseline=-0.5ex]
    \draw[midarrow] (135:0.2) -- (135:1) ;
    \draw[midarrow] (-135:1) -- (-135:0.2);
    \draw[dashed] (45:0.2) -- (45:1);
    \draw[dashed] (-45:0.2) -- (-45:1);
    \draw[photon] (90:0.2) -- (90:1);
    \draw[photon,color=white] (-90:0.3) -- (-90:1);
    \fill[white] (0:0) circle (0.22);
    \draw (0:0) circle (0.22);
    \node at (0:0) {1};
    \end{tikzpicture}
}

\newcommand{\fourpointeightd}[1]{
    \begin{tikzpicture}[scale=#1,baseline=-0.5ex]
        \draw[midarrow] (135:0.2) -- (135:1);
    \draw[midarrow] (-135:1) -- (-135:0.2);
    \draw[dashed] (45:0.2) -- (45:1);
    \draw[dashed] (-45:0.2) -- (-45:1);
    \fill[white] (0:0) circle (0.22);
    \draw (0:0) circle (0.22);
    \node at (0:0) {2};
    \end{tikzpicture}
}

\newcommand{\fivepointeightd}[1]{
    \begin{tikzpicture}[scale=#1,baseline=-0.5ex]
    \draw[midarrow] (135:0.2) -- (135:1) ;
    \draw[midarrow] (-135:1) -- (-135:0.2);
    \draw[dashed] (45:0.2) -- (45:1);
    \draw[dashed] (-45:0.2) -- (-45:1);
    \draw[photon] (90:0.2) -- (90:1);
    \draw[photon, color=white] (-90:0.3) -- (-90:1);
    \fill[white] (0:0) circle (0.22);
    \draw (0:0) circle (0.22);
    \node at (0:0) {2};
    \end{tikzpicture}
}

\newcommand{\sixpoint}[1]{
    \begin{tikzpicture}[scale=#1,baseline=-0.5ex]
    \draw[midarrow] (-150:0.2) -- (-150:1.) ;
    \draw[dashed] (-30:0.2) -- (-30:1.);
    \draw[midarrow] (-90:1.) -- (-90:0.2);
    \draw[midarrow] (90:0.2) -- (90:1.);
    \draw[midarrow] (150:1.0) -- (150:0.2);
    \draw[dashed] (30:0.2) -- (30:1.);
    \fill[white] (0:0) circle (0.22);
    \draw (0:0) circle (0.22);
    \node at (0:0) {3};
    \end{tikzpicture}
}

\newcommand{\hvpsix}[1]{
\begin{tikzpicture}[scale=#1,baseline=-0.5ex]
    \draw[midarrow] (135:0.2) -- (135:1) ;
    \draw[midarrow] (-135:1) -- (-135:0.2);
    \draw[midarrow] (45:1) -- (45:0.2);
    \draw[midarrow] (-45:0.2) -- (-45:1);
    \fill[white] (0:0) circle (0.22);
    \draw (0:0) circle (0.22);
    \node at (0:0) {4};
    \end{tikzpicture}
}

\newcommand{\hvpeight}[1]{
\begin{tikzpicture}[scale=#1,baseline=-0.5ex]
    \draw[midarrow] (135:0.2) -- (135:1) ;
    \draw[midarrow] (-135:1) -- (-135:0.2);
    \draw[midarrow] (45:1) -- (45:0.2);
    \draw[midarrow] (-45:0.2) -- (-45:1);
    \fill[white] (0:0) circle (0.22);
    \draw (0:0) circle (0.22);
    \node at (0:0) {5};
    \end{tikzpicture}
}

%% file: main.bbl
\providecommand{\href}[2]{#2}\begingroup\raggedright\begin{thebibliography}{100}

\bibitem{Aliberti:2025beg}
R.~Aliberti et~al., \emph{{The anomalous magnetic moment of the muon in the
  Standard Model: an update}},
  \href{https://doi.org/10.1016/j.physrep.2025.08.002}{\emph{Phys. Rept.}
  {\bfseries 1143} (2025) 1}
  [\href{https://arxiv.org/abs/2505.21476}{{\ttfamily 2505.21476}}].

\bibitem{Borsanyi:2020mff}
S.~Borsanyi et~al., \emph{{Leading hadronic contribution to the muon magnetic
  moment from lattice QCD}},
  \href{https://doi.org/10.1038/s41586-021-03418-1}{\emph{Nature} {\bfseries
  593} (2021) 51} [\href{https://arxiv.org/abs/2002.12347}{{\ttfamily
  2002.12347}}].

\bibitem{RBC:2023pvn}
{\scshape RBC, UKQCD} collaboration, T.~Blum et~al., \emph{{Update of Euclidean
  windows of the hadronic vacuum polarization}},
  \href{https://doi.org/10.1103/PhysRevD.108.054507}{\emph{Phys. Rev. D}
  {\bfseries 108} (2023) 054507}
  [\href{https://arxiv.org/abs/2301.08696}{{\ttfamily 2301.08696}}].

\bibitem{Djukanovic:2024cmq}
D.~Djukanovic, G.~von Hippel, S.~Kuberski, H.~B. Meyer, N.~Miller, K.~Ottnad
  et~al., \emph{{The hadronic vacuum polarization contribution to the muon
  $g-2$ at long distances}},
  \href{https://doi.org/10.1007/JHEP04(2025)098}{\emph{JHEP} {\bfseries 04}
  (2025) 098} [\href{https://arxiv.org/abs/2411.07969}{{\ttfamily
  2411.07969}}].

\bibitem{Afanasev:2023gev}
A.~Afanasev et~al., \emph{{Radiative corrections: from medium to high energy
  experiments}},
  \href{https://doi.org/10.1140/epja/s10050-024-01281-y}{\emph{Eur. Phys. J. A}
  {\bfseries 60} (2024) 91} [\href{https://arxiv.org/abs/2306.14578}{{\ttfamily
  2306.14578}}].

\bibitem{Pohl:2010zza}
R.~Pohl et~al., \emph{{The size of the proton}},
  \href{https://doi.org/10.1038/nature09250}{\emph{Nature} {\bfseries 466}
  (2010) 213}.

\bibitem{Aliberti:2024fpq}
R.~Aliberti et~al., \emph{{Radiative corrections and Monte Carlo tools for
  low-energy hadronic cross sections in $e^+ e^-$ collisions}},
  \href{https://doi.org/10.21468/SciPostPhysCommRep.9}{\emph{SciPost Phys.
  Comm. Rep.} {\bfseries 2025} (2025) 9}
  [\href{https://arxiv.org/abs/2410.22882}{{\ttfamily 2410.22882}}].

\bibitem{Bucoveanu:2018soy}
R.~D. Bucoveanu and H.~Spiesberger, \emph{{Second-Order Leptonic Radiative
  Corrections for Lepton-Proton Scattering}},
  \href{https://doi.org/10.1140/epja/i2019-12727-1}{\emph{Eur. Phys. J. A}
  {\bfseries 55} (2019) 57} [\href{https://arxiv.org/abs/1811.04970}{{\ttfamily
  1811.04970}}].

\bibitem{Banerjee:2020rww}
P.~Banerjee, T.~Engel, A.~Signer and Y.~Ulrich, \emph{{QED at NNLO with
  McMule}}, \href{https://doi.org/10.21468/SciPostPhys.9.2.027}{\emph{SciPost
  Phys.} {\bfseries 9} (2020) 027}
  [\href{https://arxiv.org/abs/2007.01654}{{\ttfamily 2007.01654}}].

\bibitem{CarloniCalame:2020yoz}
C.~M. Carloni~Calame, M.~Chiesa, S.~M. Hasan, G.~Montagna, O.~Nicrosini and
  F.~Piccinini, \emph{{Towards muon-electron scattering at NNLO}},
  \href{https://doi.org/10.1007/JHEP11(2020)028}{\emph{JHEP} {\bfseries 11}
  (2020) 028} [\href{https://arxiv.org/abs/2007.01586}{{\ttfamily
  2007.01586}}].

\bibitem{Banerjee:2021mty}
P.~Banerjee, T.~Engel, N.~Schalch, A.~Signer and Y.~Ulrich, \emph{{Bhabha
  scattering at NNLO with next-to-soft stabilisation}},
  \href{https://doi.org/10.1016/j.physletb.2021.136547}{\emph{Phys. Lett. B}
  {\bfseries 820} (2021) 136547}
  [\href{https://arxiv.org/abs/2106.07469}{{\ttfamily 2106.07469}}].

\bibitem{Banerjee:2021qvi}
P.~Banerjee, T.~Engel, N.~Schalch, A.~Signer and Y.~Ulrich, \emph{{M{\o}ller
  scattering at NNLO}},
  \href{https://doi.org/10.1103/PhysRevD.105.L031904}{\emph{Phys. Rev. D}
  {\bfseries 105} (2022) L031904}
  [\href{https://arxiv.org/abs/2107.12311}{{\ttfamily 2107.12311}}].

\bibitem{Budassi:2021twh}
E.~Budassi, C.~M. Carloni~Calame, M.~Chiesa, C.~L. Del~Pio, S.~M. Hasan,
  G.~Montagna et~al., \emph{{NNLO virtual and real leptonic corrections to
  muon-electron scattering}},
  \href{https://doi.org/10.1007/JHEP11(2021)098}{\emph{JHEP} {\bfseries 11}
  (2021) 098} [\href{https://arxiv.org/abs/2109.14606}{{\ttfamily
  2109.14606}}].

\bibitem{Broggio:2022htr}
A.~Broggio et~al., \emph{{Muon-electron scattering at NNLO}},
  \href{https://doi.org/10.1007/JHEP01(2023)112}{\emph{JHEP} {\bfseries 01}
  (2023) 112} [\href{https://arxiv.org/abs/2212.06481}{{\ttfamily
  2212.06481}}].

\bibitem{Engel:2023arz}
T.~Engel, F.~Hagelstein, M.~Rocco, V.~Sharkovska, A.~Signer and Y.~Ulrich,
  \emph{{Impact of NNLO QED corrections on lepton-proton scattering at MUSE}},
  \href{https://doi.org/10.1140/epja/s10050-023-01153-x}{\emph{Eur. Phys. J. A}
  {\bfseries 59} (2023) 253}
  [\href{https://arxiv.org/abs/2307.16831}{{\ttfamily 2307.16831}}].

\bibitem{Balossini:2006wc}
G.~Balossini, C.~M. Carloni~Calame, G.~Montagna, O.~Nicrosini and F.~Piccinini,
  \emph{{Matching perturbative and parton shower corrections to Bhabha process
  at flavour factories}},
  \href{https://doi.org/10.1016/j.nuclphysb.2006.09.022}{\emph{Nucl. Phys. B}
  {\bfseries 758} (2006) 227}
  [\href{https://arxiv.org/abs/hep-ph/0607181}{{\ttfamily hep-ph/0607181}}].

\bibitem{Balossini:2008xr}
G.~Balossini, C.~Bignamini, C.~M.~C. Calame, G.~Montagna, O.~Nicrosini and
  F.~Piccinini, \emph{{Photon pair production at flavour factories with per
  mille accuracy}},
  \href{https://doi.org/10.1016/j.physletb.2008.04.007}{\emph{Phys. Lett. B}
  {\bfseries 663} (2008) 209}
  [\href{https://arxiv.org/abs/0801.3360}{{\ttfamily 0801.3360}}].

\bibitem{Budassi:2024whw}
E.~Budassi, C.~M. Carloni~Calame, M.~Ghilardi, A.~Gurgone, G.~Montagna,
  M.~Moretti et~al., \emph{{Pion pair production in $e^+e^-$ annihilation at
  next-to-leading order matched to Parton Shower}},
  \href{https://doi.org/10.1007/JHEP05(2025)196}{\emph{JHEP} {\bfseries 05}
  (2025) 196} [\href{https://arxiv.org/abs/2409.03469}{{\ttfamily
  2409.03469}}].

\bibitem{Price:2025fiu}
A.~Price and F.~Krauss, \emph{{Towards a Fully Automated Differential
  $\text{NNLO}_\text{EW}$ Generator for Lepton Colliders}},
  \href{https://arxiv.org/abs/2512.04959}{{\ttfamily 2512.04959}}.

\bibitem{Alacevich:2018vez}
M.~Alacevich, C.~M. Carloni~Calame, M.~Chiesa, G.~Montagna, O.~Nicrosini and
  F.~Piccinini, \emph{{Muon-electron scattering at NLO}},
  \href{https://doi.org/10.1007/JHEP02(2019)155}{\emph{JHEP} {\bfseries 02}
  (2019) 155} [\href{https://arxiv.org/abs/1811.06743}{{\ttfamily
  1811.06743}}].

\bibitem{Arbuzov:2022mij}
A.~Arbuzov, S.~Bondarenko, Y.~Dydyshka, L.~Kalinovskaya, L.~Rumyantsev,
  R.~Sadykov et~al., \emph{{Effects of Electroweak Radiative Corrections in
  Polarized Low-Energy Electron{\textendash}Positron Annihilation into Lepton
  Pairs}}, \href{https://doi.org/10.1134/S0021364022601415}{\emph{JETP Lett.}
  {\bfseries 116} (2022) 199}
  [\href{https://arxiv.org/abs/2206.09469}{{\ttfamily 2206.09469}}].

\bibitem{Kollatzsch:2022bqa}
S.~Kollatzsch and Y.~Ulrich, \emph{{Lepton pair production at NNLO in QED with
  EW effects}},
  \href{https://doi.org/10.21468/SciPostPhys.15.3.104}{\emph{SciPost Phys.}
  {\bfseries 15} (2023) 104}
  [\href{https://arxiv.org/abs/2210.17172}{{\ttfamily 2210.17172}}].

\bibitem{Kollatzsch:2025pnp}
S.~Kollatzsch, D.~Moreno, D.~Radic and A.~Signer, \emph{{Parity violation in
  M{\o}ller scattering within low-energy effective field theory}},
  \href{https://doi.org/10.1007/JHEP09(2025)196}{\emph{JHEP} {\bfseries 09}
  (2025) 196} [\href{https://arxiv.org/abs/2507.17652}{{\ttfamily
  2507.17652}}].

\bibitem{Fadin:2023phc}
V.~S. Fadin and R.~N. Lee, \emph{{Two-loop radiative corrections to $e^+ e^-
  \to \gamma\gamma^{*}$ cross section}},
  \href{https://doi.org/10.1007/JHEP11(2023)148}{\emph{JHEP} {\bfseries 11}
  (2023) 148} [\href{https://arxiv.org/abs/2308.09479}{{\ttfamily
  2308.09479}}].

\bibitem{Badger:2023xtl}
S.~Badger, J.~Kry{\'s}, R.~Moodie and S.~Zoia, \emph{{Lepton-pair scattering
  with an off-shell and an on-shell photon at two loops in massless QED}},
  \href{https://doi.org/10.1007/JHEP11(2023)041}{\emph{JHEP} {\bfseries 11}
  (2023) 041} [\href{https://arxiv.org/abs/2307.03098}{{\ttfamily
  2307.03098}}].

\bibitem{Badger:2023mgf}
S.~Badger, M.~Czakon, H.~B. Hartanto, R.~Moodie, T.~Peraro, R.~Poncelet et~al.,
  \emph{{Isolated photon production in association with a jet pair through
  next-to-next-to-leading order in QCD}},
  \href{https://doi.org/10.1007/JHEP10(2023)071}{\emph{JHEP} {\bfseries 10}
  (2023) 071} [\href{https://arxiv.org/abs/2304.06682}{{\ttfamily
  2304.06682}}].

\bibitem{PetitRosas:2025xhm}
P.~Petit~Ros{\`a}s and W.~J. Torres~Bobadilla, \emph{{Fast evaluation of
  Feynman integrals for Monte Carlo generators}},
  \href{https://doi.org/10.1007/JHEP09(2025)210}{\emph{JHEP} {\bfseries 09}
  (2025) 210} [\href{https://arxiv.org/abs/2507.12548}{{\ttfamily
  2507.12548}}].

\bibitem{Hoferichter:2013ama}
M.~Hoferichter, G.~Colangelo, M.~Procura and P.~Stoffer, \emph{{Virtual
  photon-photon scattering}},
  \href{https://doi.org/10.1142/S2010194514604001}{\emph{Int. J. Mod. Phys.
  Conf. Ser.} {\bfseries 35} (2014) 1460400}
  [\href{https://arxiv.org/abs/1309.6877}{{\ttfamily 1309.6877}}].

\bibitem{Colangelo:2015ama}
G.~Colangelo, M.~Hoferichter, M.~Procura and P.~Stoffer, \emph{{Dispersion
  relation for hadronic light-by-light scattering: theoretical foundations}},
  \href{https://doi.org/10.1007/JHEP09(2015)074}{\emph{JHEP} {\bfseries 09}
  (2015) 074} [\href{https://arxiv.org/abs/1506.01386}{{\ttfamily
  1506.01386}}].

\bibitem{Hoferichter:2019nlq}
M.~Hoferichter and P.~Stoffer, \emph{{Dispersion relations for
  $\gamma^*\gamma^*\to\pi\pi$: helicity amplitudes, subtractions, and anomalous
  thresholds}}, \href{https://doi.org/10.1007/JHEP07(2019)073}{\emph{JHEP}
  {\bfseries 07} (2019) 073}
  [\href{https://arxiv.org/abs/1905.13198}{{\ttfamily 1905.13198}}].

\bibitem{Mo:1968cg}
L.~W. Mo and Y.-S. Tsai, \emph{{Radiative Corrections to Elastic and Inelastic
  $e p$ and $\mu p$ Scattering}},
  \href{https://doi.org/10.1103/RevModPhys.41.205}{\emph{Rev. Mod. Phys.}
  {\bfseries 41} (1969) 205}.

\bibitem{Maximon:2000hm}
L.~C. Maximon and J.~A. Tjon, \emph{{Radiative corrections to electron proton
  scattering}}, \href{https://doi.org/10.1103/PhysRevC.62.054320}{\emph{Phys.
  Rev. C} {\bfseries 62} (2000) 054320}.

\bibitem{Kaziukenas:2025gggpp}
{Emilis Kaziuk{\.e}nas}, \emph{{Dispersive definition of
  $\gamma^*\gamma^*\gamma\to\pi^+\pi^-$ for muon $g-2$ applications}},  in
  \emph{{RadioMonteCarLow 2 Satellite 2025}}, (Liverpool), 2025,
  \href{https://indico.ph.liv.ac.uk/event/2169/contributions/10215/}{https://indico.ph.liv.ac.uk/event/2169/contributions/10215/}.

\bibitem{Fael:2019nsf}
M.~Fael and M.~Passera, \emph{{Muon-Electron Scattering at
  Next-To-Next-To-Leading Order: The Hadronic Corrections}},
  \href{https://doi.org/10.1103/PhysRevLett.122.192001}{\emph{Phys. Rev. Lett.}
  {\bfseries 122} (2019) 192001}
  [\href{https://arxiv.org/abs/1901.03106}{{\ttfamily 1901.03106}}].

\bibitem{Fael:2018dmz}
M.~Fael, \emph{{Hadronic corrections to $\mu$-$e$ scattering at NNLO with
  space-like data}}, \href{https://doi.org/10.1007/JHEP02(2019)027}{\emph{JHEP}
  {\bfseries 02} (2019) 027}
  [\href{https://arxiv.org/abs/1808.08233}{{\ttfamily 1808.08233}}].

\bibitem{Balzani:2021del}
E.~Balzani, S.~Laporta and M.~Passera, \emph{{Hadronic vacuum polarization
  contributions to the muon $g-2$ in the space-like region}},
  \href{https://doi.org/10.1016/j.physletb.2022.137462}{\emph{Phys. Lett. B}
  {\bfseries 834} (2022) 137462}
  [\href{https://arxiv.org/abs/2112.05704}{{\ttfamily 2112.05704}}].

\bibitem{Sakurai:1972wk}
J.~J. Sakurai and D.~Schildknecht, \emph{{Generalized vector dominance and
  inelastic electron - proton scattering}},
  \href{https://doi.org/10.1016/0370-2693(72)90300-0}{\emph{Phys. Lett. B}
  {\bfseries 40} (1972) 121}.

\bibitem{Ignatov:2022iou}
F.~Ignatov and R.~N. Lee, \emph{{Charge asymmetry in $e^+e^-\to\pi^+\pi^-$
  process}}, \href{https://doi.org/10.1016/j.physletb.2022.137283}{\emph{Phys.
  Lett. B} {\bfseries 833} (2022) 137283}
  [\href{https://arxiv.org/abs/2204.12235}{{\ttfamily 2204.12235}}].

\bibitem{Gramolin:2014pva}
A.~V. Gramolin, V.~S. Fadin, A.~L. Feldman, R.~E. Gerasimov, D.~M. Nikolenko,
  I.~A. Rachek et~al., \emph{{A new event generator for the elastic scattering
  of charged leptons on protons}},
  \href{https://doi.org/10.1088/0954-3899/41/11/115001}{\emph{J. Phys.}
  {\bfseries G41} (2014) 115001}.

\bibitem{Gerasimov:2015aoa}
R.~E. Gerasimov and V.~S. Fadin, \emph{{Analysis of approximations used in
  calculations of radiative corrections to electron-proton scattering cross
  section}}, \href{https://doi.org/10.1134/S1063778815010081}{\emph{Phys. Atom.
  Nucl.} {\bfseries 78} (2015) 69}.

\bibitem{Bystritskiy:2007hw}
Y.~Bystritskiy, E.~Kuraev and E.~Tomasi-Gustafsson, \emph{{Structure function
  method applied to polarized and unpolarized electron-proton scattering: A
  solution of the $G_E(p)$/$G_M(p)$ discrepancy}},
  \href{https://doi.org/10.1103/PhysRevC.75.015207}{\emph{Phys. Rev.}
  {\bfseries C75} (2007) 015207}.

\bibitem{Choudhary:2023rsz}
P.~Choudhary, U.~Raha, F.~Myhrer and D.~Chakrabarti, \emph{{Analytical
  evaluation of elastic lepton-proton two-photon exchange in chiral
  perturbation theory}},
  \href{https://doi.org/10.1140/epja/s10050-023-01207-0}{\emph{Eur. Phys. J. A}
  {\bfseries 60} (2024) 69} [\href{https://arxiv.org/abs/2306.09454}{{\ttfamily
  2306.09454}}].

\bibitem{Dye:2016uep}
S.~P. Dye, M.~Gonderinger and G.~Paz, \emph{{Elements of QED-NRQED effective
  field theory: NLO scattering at leading power}},
  \href{https://doi.org/10.1103/PhysRevD.94.013006}{\emph{Phys. Rev. D}
  {\bfseries 94} (2016) 013006}
  [\href{https://arxiv.org/abs/1602.07770}{{\ttfamily 1602.07770}}].

\bibitem{Dye:2018rgg}
S.~P. Dye, M.~Gonderinger and G.~Paz, \emph{{Elements of QED-NRQED Effective
  Field Theory: II. Matching of Contact Interactions}},
  \href{https://doi.org/10.1103/PhysRevD.100.054010}{\emph{Phys. Rev. D}
  {\bfseries 100} (2019) 054010}
  [\href{https://arxiv.org/abs/1812.05056}{{\ttfamily 1812.05056}}].

\bibitem{Tomalak:2016vbf}
O.~Tomalak, B.~Pasquini and M.~Vanderhaeghen, \emph{{Two-photon exchange
  corrections to elastic $e^-$-proton scattering: Full dispersive treatment of
  $\pi N$ states at low momentum transfers}},
  \href{https://doi.org/10.1103/PhysRevD.95.096001}{\emph{Phys. Rev. D}
  {\bfseries 95} (2017) 096001}
  [\href{https://arxiv.org/abs/1612.07726}{{\ttfamily 1612.07726}}].

\bibitem{Tomalak:2017shs}
O.~Tomalak, B.~Pasquini and M.~Vanderhaeghen, \emph{{Two-photon exchange
  contribution to elastic $e^-$ -proton scattering: Full dispersive treatment
  of $\pi N$ states and comparison with data}},
  \href{https://doi.org/10.1103/PhysRevD.96.096001}{\emph{Phys. Rev. D}
  {\bfseries 96} (2017) 096001}
  [\href{https://arxiv.org/abs/1708.03303}{{\ttfamily 1708.03303}}].

\bibitem{Ahmed:2020uso}
J.~Ahmed, P.~G. Blunden and W.~Melnitchouk, \emph{{Two-photon exchange from
  intermediate state resonances in elastic electron-proton scattering}},
  \href{https://doi.org/10.1103/PhysRevC.102.045205}{\emph{Phys. Rev. C}
  {\bfseries 102} (2020) 045205}
  [\href{https://arxiv.org/abs/2006.12543}{{\ttfamily 2006.12543}}].

\bibitem{Carlson:2007sp}
C.~E. Carlson and M.~Vanderhaeghen, \emph{{Two-Photon Physics in Hadronic
  Processes}},
  \href{https://doi.org/10.1146/annurev.nucl.57.090506.123116}{\emph{Ann. Rev.
  Nucl. Part. Sci.} {\bfseries 57} (2007) 171}
  [\href{https://arxiv.org/abs/hep-ph/0701272}{{\ttfamily hep-ph/0701272}}].

\bibitem{Arrington:2011dn}
J.~Arrington, P.~G. Blunden and W.~Melnitchouk, \emph{{Review of two-photon
  exchange in electron scattering}},
  \href{https://doi.org/10.1016/j.ppnp.2011.07.003}{\emph{Prog. Part. Nucl.
  Phys.} {\bfseries 66} (2011) 782}
  [\href{https://arxiv.org/abs/1105.0951}{{\ttfamily 1105.0951}}].

\bibitem{Afanasev:2017gsk}
A.~Afanasev, P.~G. Blunden, D.~Hasell and B.~A. Raue, \emph{{Two-photon
  exchange in elastic electron\textendash{}proton scattering}},
  \href{https://doi.org/10.1016/j.ppnp.2017.03.004}{\emph{Prog. Part. Nucl.
  Phys.} {\bfseries 95} (2017) 245}
  [\href{https://arxiv.org/abs/1703.03874}{{\ttfamily 1703.03874}}].

\bibitem{Borisyuk:2019gym}
D.~Borisyuk and A.~Kobushkin, \emph{{Two-Photon Exchange in Elastic Electron
  Scattering on Hadronic Systems}},
  \href{https://doi.org/10.15407/ujpe66.1.3}{\emph{Ukr. J. Phys.} {\bfseries
  66} (2021) 3} [\href{https://arxiv.org/abs/1911.10956}{{\ttfamily
  1911.10956}}].

\bibitem{Cabibbo:1961sz}
N.~Cabibbo and R.~Gatto, \emph{{Electron Positron Colliding Beam Experiments}},
  \href{https://doi.org/10.1103/PhysRev.124.1577}{\emph{Phys. Rev.} {\bfseries
  124} (1961) 1577}.

\bibitem{Colangelo:2022lzg}
G.~Colangelo, M.~Hoferichter, J.~Monnard and J.~R. de~Elvira, \emph{{Radiative
  corrections to the forward-backward asymmetry in $e^+e^-\to\pi^+\pi^-$}},
  \href{https://doi.org/10.1007/JHEP08(2022)295}{\emph{JHEP} {\bfseries 08}
  (2022) 295} [\href{https://arxiv.org/abs/2207.03495}{{\ttfamily
  2207.03495}}].

\bibitem{CMD-3:2023alj}
{\scshape CMD-3} collaboration, F.~V. Ignatov et~al., \emph{{Measurement of the
  $e^+e^-\to\pi^+\pi^-$ cross section from threshold to 1.2~GeV with the CMD-3
  detector}}, \href{https://doi.org/10.1103/PhysRevD.109.112002}{\emph{Phys.
  Rev. D} {\bfseries 109} (2024) 112002}
  [\href{https://arxiv.org/abs/2302.08834}{{\ttfamily 2302.08834}}].

\bibitem{CMD-3:2023rfe}
{\scshape CMD-3} collaboration, F.~V. Ignatov et~al., \emph{{Measurement of the
  Pion Form Factor with CMD-3 Detector and its Implication to the Hadronic
  Contribution to Muon $g-2$}},
  \href{https://doi.org/10.1103/PhysRevLett.132.231903}{\emph{Phys. Rev. Lett.}
  {\bfseries 132} (2024) 231903}
  [\href{https://arxiv.org/abs/2309.12910}{{\ttfamily 2309.12910}}].

\bibitem{Hoefer:2001mx}
A.~Hoefer, J.~Gluza and F.~Jegerlehner, \emph{{Pion pair production with higher
  order radiative corrections in low energy $e^+ e^-$ collisions}},
  \href{https://doi.org/10.1007/s100520200916}{\emph{Eur. Phys. J. C}
  {\bfseries 24} (2002) 51}
  [\href{https://arxiv.org/abs/hep-ph/0107154}{{\ttfamily hep-ph/0107154}}].

\bibitem{Tracz:2018}
S.~Tracz, \emph{{Radiative corrections to hadrons-photons interactions}}, Ph.D.
  thesis, University of Silesia, 2018.

\bibitem{Arbuzov:2020foj}
A.~B. Arbuzov, T.~V. Kopylova and G.~A. Seilkhanova,
  \emph{{Forward{\textendash}backward asymmetry in
  electron{\textendash}positron annihilation into pion or kaon pairs
  revisited}}, \href{https://doi.org/10.1142/S0217732320502107}{\emph{Mod.
  Phys. Lett. A} {\bfseries 35} (2020) 2050210}
  [\href{https://arxiv.org/abs/2003.14054}{{\ttfamily 2003.14054}}].

\bibitem{Colangelo:2014dfa}
G.~Colangelo, M.~Hoferichter, M.~Procura and P.~Stoffer, \emph{{Dispersive
  approach to hadronic light-by-light scattering}},
  \href{https://doi.org/10.1007/JHEP09(2014)091}{\emph{JHEP} {\bfseries 09}
  (2014) 091} [\href{https://arxiv.org/abs/1402.7081}{{\ttfamily 1402.7081}}].

\bibitem{Czyz:2003ue}
H.~Czy{\.z}, A.~Grzelinska, J.~H. Kuhn and G.~Rodrigo, \emph{{The Radiative
  return at $\Phi$ and B factories: FSR at next-to-leading order}},
  \href{https://doi.org/10.1140/epjc/s2004-01605-0}{\emph{Eur. Phys. J. C}
  {\bfseries 33} (2004) 333}
  [\href{https://arxiv.org/abs/hep-ph/0308312}{{\ttfamily hep-ph/0308312}}].

\bibitem{Campanario:2019mjh}
F.~Campanario, H.~Czy{\.z}, J.~Gluza, T.~Jeli{\'n}ski, G.~Rodrigo, S.~Tracz
  et~al., \emph{{Standard model radiative corrections in the pion form factor
  measurements do not explain the $a_\mu$ anomaly}},
  \href{https://doi.org/10.1103/PhysRevD.100.076004}{\emph{Phys. Rev. D}
  {\bfseries 100} (2019) 076004}
  [\href{https://arxiv.org/abs/1903.10197}{{\ttfamily 1903.10197}}].

\bibitem{McMule:manual}
\mcmule{} Team, ``\mcmule{} manual.''
  \url{https://doi.org/10.5281/zenodo.6046769}.

\bibitem{Engel:2019nfw}
T.~Engel, A.~Signer and Y.~Ulrich, \emph{{A subtraction scheme for massive
  QED}}, \href{https://doi.org/10.1007/JHEP01(2020)085}{\emph{JHEP} {\bfseries
  01} (2020) 085} [\href{https://arxiv.org/abs/1909.10244}{{\ttfamily
  1909.10244}}].

\bibitem{Buccioni:2017yxi}
F.~Buccioni, S.~Pozzorini and M.~Zoller, \emph{{On-the-fly reduction of open
  loops}}, \href{https://doi.org/10.1140/epjc/s10052-018-5562-1}{\emph{Eur.
  Phys. J. C} {\bfseries 78} (2018) 70}
  [\href{https://arxiv.org/abs/1710.11452}{{\ttfamily 1710.11452}}].

\bibitem{Buccioni:2019sur}
F.~Buccioni, J.-N. Lang, J.~M. Lindert, P.~Maierh{\"{o}}fer, S.~Pozzorini,
  H.~Zhang et~al., \emph{{OpenLoops 2}},
  \href{https://doi.org/10.1140/epjc/s10052-019-7306-2}{\emph{Eur. Phys. J. C}
  {\bfseries 79} (2019) 866}
  [\href{https://arxiv.org/abs/1907.13071}{{\ttfamily 1907.13071}}].

\bibitem{Jegerlehner:2017gek}
F.~Jegerlehner, \emph{{The Anomalous Magnetic Moment of the Muon}}, vol.~274.
  Springer, Cham, 2017,
  \href{https://doi.org/10.1007/978-3-319-63577-4}{10.1007/978-3-319-63577-4}.

\bibitem{Patel:2015tea}
H.~H. Patel, \emph{{Package-X: A Mathematica package for the analytic
  calculation of one-loop integrals}},
  \href{https://doi.org/10.1016/j.cpc.2015.08.017}{\emph{Comput. Phys. Commun.}
  {\bfseries 197} (2015) 276}
  [\href{https://arxiv.org/abs/1503.01469}{{\ttfamily 1503.01469}}].

\bibitem{Patel:2016fam}
H.~H. Patel, \emph{{Package-X 2.0: A Mathematica package for the analytic
  calculation of one-loop integrals}},
  \href{https://doi.org/10.1016/j.cpc.2017.04.015}{\emph{Comput. Phys. Commun.}
  {\bfseries 218} (2017) 66}
  [\href{https://arxiv.org/abs/1612.00009}{{\ttfamily 1612.00009}}].

\bibitem{Hoferichter:2025rescatteringpp}
{Martin Hoferichter}, \emph{{Rescattering corrections to the pion Compton
  tensor }},  in \emph{{RadioMonteCarLow 2 Satellite 2025}}, (Liverpool), 2025,
  \href{https://indico.ph.liv.ac.uk/event/2169/contributions/10214/}{https://indico.ph.liv.ac.uk/event/2169/contributions/10214/}.

\bibitem{Monnard:2021pvm}
J.~Monnard, \emph{{Radiative corrections for the two-pion contribution to the
  hadronic vacuum polarization contribution to the muon $g-2$}}, Ph.D. thesis,
  Bern U., 2021.

\bibitem{Farrar:1979aw}
G.~R. Farrar and D.~R. Jackson, \emph{{The Pion Form-Factor}},
  \href{https://doi.org/10.1103/PhysRevLett.43.246}{\emph{Phys. Rev. Lett.}
  {\bfseries 43} (1979) 246}.

\bibitem{tHooft:1972tcz}
G.~'t~Hooft and M.~J.~G. Veltman, \emph{{Regularization and Renormalization of
  Gauge Fields}},
  \href{https://doi.org/10.1016/0550-3213(72)90279-9}{\emph{Nucl. Phys. B}
  {\bfseries 44} (1972) 189}.

\bibitem{Ossola:2008xq}
G.~Ossola, C.~G. Papadopoulos and R.~Pittau, \emph{{On the Rational Terms of
  the one-loop amplitudes}},
  \href{https://doi.org/10.1088/1126-6708/2008/05/004}{\emph{JHEP} {\bfseries
  05} (2008) 004} [\href{https://arxiv.org/abs/0802.1876}{{\ttfamily
  0802.1876}}].

\bibitem{Garzelli:2009is}
M.~V. Garzelli, I.~Malamos and R.~Pittau, \emph{{Feynman rules for the rational
  part of the Electroweak 1-loop amplitudes}},
  \href{https://doi.org/10.1007/JHEP10(2010)097}{\emph{JHEP} {\bfseries 01}
  (2010) 040} [\href{https://arxiv.org/abs/0910.3130}{{\ttfamily 0910.3130}}].

\bibitem{Bern:2002zk}
Z.~Bern, A.~De~Freitas, L.~J. Dixon and H.~L. Wong, \emph{{Supersymmetric
  regularization, two loop QCD amplitudes and coupling shifts}},
  \href{https://doi.org/10.1103/PhysRevD.66.085002}{\emph{Phys. Rev. D}
  {\bfseries 66} (2002) 085002}
  [\href{https://arxiv.org/abs/hep-ph/0202271}{{\ttfamily hep-ph/0202271}}].

\bibitem{Kilgore:2012tb}
W.~B. Kilgore, \emph{{The Four Dimensional Helicity Scheme Beyond One Loop}},
  \href{https://doi.org/10.1103/PhysRevD.86.014019}{\emph{Phys. Rev. D}
  {\bfseries 86} (2012) 014019}
  [\href{https://arxiv.org/abs/1205.4015}{{\ttfamily 1205.4015}}].

\bibitem{Gnendiger:2014nxa}
C.~Gnendiger, A.~Signer and D.~St{\"o}ckinger, \emph{{The infrared structure of
  QCD amplitudes and $H \to gg$ in FDH and DRED}},
  \href{https://doi.org/10.1016/j.physletb.2014.05.003}{\emph{Phys. Lett. B}
  {\bfseries 733} (2014) 296}
  [\href{https://arxiv.org/abs/1404.2171}{{\ttfamily 1404.2171}}].

\bibitem{Gnendiger:2017pys}
C.~Gnendiger et~al., \emph{{To ${d}$, or not to ${d}$: recent developments and
  comparisons of regularization schemes}},
  \href{https://doi.org/10.1140/epjc/s10052-017-5023-2}{\emph{Eur. Phys. J. C}
  {\bfseries 77} (2017) 471}
  [\href{https://arxiv.org/abs/1705.01827}{{\ttfamily 1705.01827}}].

\bibitem{Beneke:1997zp}
M.~Beneke and V.~A. Smirnov, \emph{{Asymptotic expansion of Feynman integrals
  near threshold}},
  \href{https://doi.org/10.1016/S0550-3213(98)00138-2}{\emph{Nucl. Phys. B}
  {\bfseries 522} (1998) 321}
  [\href{https://arxiv.org/abs/hep-ph/9711391}{{\ttfamily hep-ph/9711391}}].

\bibitem{Nogueira:1991ex}
P.~Nogueira, \emph{{Automatic Feynman Graph Generation}},
  \href{https://doi.org/10.1006/jcph.1993.1074}{\emph{J. Comput. Phys.}
  {\bfseries 105} (1993) 279}.

\bibitem{McMule:data}
\mcmule{} Team, ``\mcmule{} dataset.''
  \url{https://doi.org/10.5281/zenodo.8188752}.

\bibitem{Kollatzsch:2026ubi}
S.~Kollatzsch, \emph{{Beyond QED: Electroweak and hadronic extensions of
  McMule}},  in \emph{{17th International Symposium on Radiative Corrections:
  Applications of Quantum Field Theory to Phenomenolog}}, 3, 2026,
  \href{https://arxiv.org/abs/2603.09443}{{\ttfamily 2603.09443}}.

\bibitem{Kuhn:2008zs}
J.~H. K{\"u}hn and S.~Uccirati, \emph{{Two-loop QED hadronic corrections to
  Bhabha scattering}},
  \href{https://doi.org/10.1016/j.nuclphysb.2008.08.002}{\emph{Nucl. Phys. B}
  {\bfseries 806} (2009) 300}
  [\href{https://arxiv.org/abs/0807.1284}{{\ttfamily 0807.1284}}].

\bibitem{Cutkosky:1960sp}
R.~E. Cutkosky, \emph{{Singularities and discontinuities of Feynman
  amplitudes}}, \href{https://doi.org/10.1063/1.1703676}{\emph{J. Math. Phys.}
  {\bfseries 1} (1960) 429}.

\bibitem{Landau:1959fi}
L.~D. Landau, \emph{{On the Analytic Properties of Vertex Parts in Quantum
  Field Theory}},
  \href{https://doi.org/10.1016/B978-0-08-010586-4.50103-6}{\emph{Zh. Eksp.
  Teor. Fiz.} {\bfseries 37} (1960) 62}.

\bibitem{Eden:1966dnq}
R.~J. Eden, P.~V. Landshoff, D.~I. Olive and J.~C. Polkinghorne, \emph{{The
  analytic S-matrix}}. Cambridge Univ. Press, Cambridge, 1966.

\bibitem{Fevola:2023kaw}
C.~Fevola, S.~Mizera and S.~Telen, \emph{{Landau Singularities Revisited:
  Computational Algebraic Geometry for Feynman Integrals}},
  \href{https://doi.org/10.1103/PhysRevLett.132.101601}{\emph{Phys. Rev. Lett.}
  {\bfseries 132} (2024) 101601}
  [\href{https://arxiv.org/abs/2311.14669}{{\ttfamily 2311.14669}}].

\bibitem{Fevola:2023fzn}
C.~Fevola, S.~Mizera and S.~Telen, \emph{{Principal Landau determinants}},
  \href{https://doi.org/10.1016/j.cpc.2024.109278}{\emph{Comput. Phys. Commun.}
  {\bfseries 303} (2024) 109278}
  [\href{https://arxiv.org/abs/2311.16219}{{\ttfamily 2311.16219}}].

\bibitem{PLDwrapper}
``{PLD-Wrapper}.'' \url{https://github.com/Tracque/PLD-Wrapper}.

\bibitem{Pak:2010pt}
A.~Pak and A.~Smirnov, \emph{{Geometric approach to asymptotic expansion of
  Feynman integrals}},
  \href{https://doi.org/10.1140/epjc/s10052-011-1626-1}{\emph{Eur. Phys. J. C}
  {\bfseries 71} (2011) 1626}
  [\href{https://arxiv.org/abs/1011.4863}{{\ttfamily 1011.4863}}].

\bibitem{Jantzen:2012mw}
B.~Jantzen, A.~V. Smirnov and V.~A. Smirnov, \emph{{Expansion by regions:
  revealing potential and Glauber regions automatically}},
  \href{https://doi.org/10.1140/epjc/s10052-012-2139-2}{\emph{Eur. Phys. J. C}
  {\bfseries 72} (2012) 2139}
  [\href{https://arxiv.org/abs/1206.0546}{{\ttfamily 1206.0546}}].

\bibitem{Heinrich:2021dbf}
G.~Heinrich, S.~Jahn, S.~P. Jones, M.~Kerner, F.~Langer, V.~Magerya et~al.,
  \emph{{Expansion by regions with pySecDec}},
  \href{https://doi.org/10.1016/j.cpc.2021.108267}{\emph{Comput. Phys. Commun.}
  {\bfseries 273} (2022) 108267}
  [\href{https://arxiv.org/abs/2108.10807}{{\ttfamily 2108.10807}}].

\bibitem{Huber:2005yg}
T.~Huber and D.~Ma{\^\i}tre, \emph{{HypExp, a Mathematica package for expanding
  hypergeometric functions around integer-valued parameters}},
  \href{https://doi.org/10.1016/j.cpc.2006.01.007}{\emph{Comput. Phys. Commun.}
  {\bfseries 175} (2006) 122}
  [\href{https://arxiv.org/abs/hep-ph/0507094}{{\ttfamily hep-ph/0507094}}].

\bibitem{Huber:2007dx}
T.~Huber and D.~Ma{\^\i}tre, \emph{{HypExp 2, Expanding hypergeometric
  functions about half-integer parameters}},
  \href{https://doi.org/10.1016/j.cpc.2007.12.008}{\emph{Comput. Phys. Commun.}
  {\bfseries 178} (2008) 755}
  [\href{https://arxiv.org/abs/0708.2443}{{\ttfamily 0708.2443}}].

\bibitem{Yennie:1961ad}
D.~R. Yennie, S.~C. Frautschi and H.~Suura, \emph{{The infrared divergence
  phenomena and high-energy processes}},
  \href{https://doi.org/10.1016/0003-4916(61)90151-8}{\emph{Annals Phys.}
  {\bfseries 13} (1961) 379}.

\bibitem{Denner:2002ii}
A.~Denner and S.~Dittmaier, \emph{{Reduction of one loop tensor five point
  integrals}}, \href{https://doi.org/10.1016/S0550-3213(03)00184-6}{\emph{Nucl.
  Phys. B} {\bfseries 658} (2003) 175}
  [\href{https://arxiv.org/abs/hep-ph/0212259}{{\ttfamily hep-ph/0212259}}].

\bibitem{Bern:1992em}
Z.~Bern, L.~J. Dixon and D.~A. Kosower, \emph{{Dimensionally regulated one loop
  integrals}}, \href{https://doi.org/10.1016/0370-2693(93)90400-C}{\emph{Phys.
  Lett. B} {\bfseries 302} (1993) 299}
  [\href{https://arxiv.org/abs/hep-ph/9212308}{{\ttfamily hep-ph/9212308}}].

\bibitem{Bern:1993kr}
Z.~Bern, L.~J. Dixon and D.~A. Kosower, \emph{{Dimensionally regulated pentagon
  integrals}}, \href{https://doi.org/10.1016/0550-3213(94)90398-0}{\emph{Nucl.
  Phys. B} {\bfseries 412} (1994) 751}
  [\href{https://arxiv.org/abs/hep-ph/9306240}{{\ttfamily hep-ph/9306240}}].

\bibitem{Colangelo:2025iad}
G.~Colangelo, M.~Cottini, M.~Hoferichter and S.~Holz, \emph{{Improved
  Calculation of Radiative Corrections to
  {\ensuremath{\tau}}{\textrightarrow}{\ensuremath{\pi}}{\ensuremath{\pi}}{\ensuremath{\nu}}{\ensuremath{\tau}}
  Decays}}, \href{https://doi.org/10.1103/ryk1-x6v1}{\emph{Phys. Rev. Lett.}
  {\bfseries 136} (2026) 101903}
  [\href{https://arxiv.org/abs/2510.26871}{{\ttfamily 2510.26871}}].

\bibitem{Colangelo:2025ivq}
G.~Colangelo, M.~Cottini, M.~Hoferichter and S.~Holz, \emph{{Radiative
  corrections to $\tau\to\pi\pi\nu_\tau$}},
  \href{https://doi.org/10.1007/JHEP02(2026)181}{\emph{JHEP} {\bfseries 02}
  (2026) 181} [\href{https://arxiv.org/abs/2511.07507}{{\ttfamily
  2511.07507}}].

\bibitem{Colangelo:2018mtw}
G.~Colangelo, M.~Hoferichter and P.~Stoffer, \emph{{Two-pion contribution to
  hadronic vacuum polarization}},
  \href{https://doi.org/10.1007/JHEP02(2019)006}{\emph{JHEP} {\bfseries 02}
  (2019) 006} [\href{https://arxiv.org/abs/1810.00007}{{\ttfamily
  1810.00007}}].

\bibitem{Colangelo:2025sah}
G.~Colangelo, \emph{{A brief introduction to dispersive methods}},  9, 2025,
  \href{https://arxiv.org/abs/2509.24548}{{\ttfamily 2509.24548}}.

\bibitem{Kniehl:1996rh}
B.~A. Kniehl, \emph{{Dispersion relations in loop calculations}}, {\emph{Acta
  Phys. Polon. B} {\bfseries 27} (1996) 3631}
  [\href{https://arxiv.org/abs/hep-ph/9607255}{{\ttfamily hep-ph/9607255}}].

\bibitem{Colangelo:2017fiz}
G.~Colangelo, M.~Hoferichter, M.~Procura and P.~Stoffer, \emph{{Dispersion
  relation for hadronic light-by-light scattering: two-pion contributions}},
  \href{https://doi.org/10.1007/JHEP04(2017)161}{\emph{JHEP} {\bfseries 04}
  (2017) 161} [\href{https://arxiv.org/abs/1702.07347}{{\ttfamily
  1702.07347}}].

\bibitem{Rocco:2025TI}
{Marco Rocco for the RadioMonteCarLow 2 Group}, \emph{{RadioMonteCarLow 2 :: An
  Overview}},  in \emph{{8th Plenary workshop of the Muon $g-2$ Theory
  Initiative 2025}}, (Orsay), 2025,
  \href{https://indico.ijclab.in2p3.fr/event/11652/contributions/39060}{https://indico.ijclab.in2p3.fr/event/11652/contributions/39060}.

\bibitem{Eichmann:2018ytt}
G.~Eichmann and G.~Ramalho, \emph{{Nucleon resonances in Compton scattering}},
  \href{https://doi.org/10.1103/PhysRevD.98.093007}{\emph{Phys. Rev. D}
  {\bfseries 98} (2018) 093007}
  [\href{https://arxiv.org/abs/1806.04579}{{\ttfamily 1806.04579}}].

\bibitem{Hagelstein:2017cbl}
F.~Hagelstein, \emph{{Exciting Nucleons in Compton Scattering and Hydrogen-Like
  Atoms}}, Ph.D. thesis, Mainz U., 2017.
\newblock \href{https://arxiv.org/abs/1710.00874}{{\ttfamily 1710.00874}}.
\newblock 10.13140/RG.2.2.25062.73281.

\bibitem{Scherer:1996ux}
S.~Scherer, A.~Y. Korchin and J.~H. Koch, \emph{{Virtual Compton scattering off
  the nucleon at low-energies}},
  \href{https://doi.org/10.1103/PhysRevC.54.904}{\emph{Phys. Rev. C} {\bfseries
  54} (1996) 904} [\href{https://arxiv.org/abs/nucl-th/9605030}{{\ttfamily
  nucl-th/9605030}}].

\bibitem{Drechsel:1997xv}
D.~Drechsel, G.~Knochlein, A.~Y. Korchin, A.~Metz and S.~Scherer,
  \emph{{Structure analysis of the virtual Compton scattering amplitude at
  low-energies}}, \href{https://doi.org/10.1103/PhysRevC.57.941}{\emph{Phys.
  Rev. C} {\bfseries 57} (1998) 941}
  [\href{https://arxiv.org/abs/nucl-th/9704064}{{\ttfamily nucl-th/9704064}}].

\bibitem{Birse:2012eb}
M.~C. Birse and J.~A. McGovern, \emph{{Proton polarisability contribution to
  the Lamb shift in muonic hydrogen at fourth order in chiral perturbation
  theory}}, \href{https://doi.org/10.1140/epja/i2012-12120-8}{\emph{Eur. Phys.
  J. A} {\bfseries 48} (2012) 120}
  [\href{https://arxiv.org/abs/1206.3030}{{\ttfamily 1206.3030}}].

\bibitem{Gasser:2015dwa}
J.~Gasser, M.~Hoferichter, H.~Leutwyler and A.~Rusetsky, \emph{{Cottingham
  formula and nucleon polarisabilities}},
  \href{https://doi.org/10.1140/epjc/s10052-015-3580-9}{\emph{Eur. Phys. J. C}
  {\bfseries 75} (2015) 375}
  [\href{https://arxiv.org/abs/1506.06747}{{\ttfamily 1506.06747}}].

\bibitem{Hagelstein:2015egb}
F.~Hagelstein, R.~Miskimen and V.~Pascalutsa, \emph{{Nucleon Polarizabilities:
  from Compton Scattering to Hydrogen Atom}},
  \href{https://doi.org/10.1016/j.ppnp.2015.12.001}{\emph{Prog. Part. Nucl.
  Phys.} {\bfseries 88} (2016) 29}
  [\href{https://arxiv.org/abs/1512.03765}{{\ttfamily 1512.03765}}].

\bibitem{Hagelstein:2018bdi}
F.~Hagelstein, \emph{{$\Delta(1232)$-Resonance in the Hydrogen Spectrum}},
  \href{https://doi.org/10.1007/s00601-018-1403-x}{\emph{Few Body Syst.}
  {\bfseries 59} (2018) 93} [\href{https://arxiv.org/abs/1801.09790}{{\ttfamily
  1801.09790}}].

\end{thebibliography}\endgroup
